\documentclass{aa} %[referee]
\usepackage{graphicx}
\usepackage{txfonts}
\usepackage[acronym]{glossaries}
\usepackage{hyperref}
\usepackage{mathtools}
\usepackage[permil]{overpic}
\usepackage{placeins} %floatbarrier
\usepackage{subcaption}
\newacronym{ccf}{CCF}{cross-correlation function}
\newacronym{drs}{DRS}{data reduction software}
\newacronym{gp}{GP}{giant planet}
\newacronym{rv}{RV}{radial-velocity}
\newacronym{di}{DI}{direct imaging}
\newacronym{ppd}{PPD}{protoplanetary disk}
\newacronym{pv}{PV}{pooled variance}
\newacronym{pvd}{PVD}{Pooled Variance Diagram}
\newacronym{sb}{SB}{spectroscopic binary}
\newacronym{mc}{MC}{Monte Carlo}
\newacronym{mcmc}{MCMC}{Markov Chain Monte Carlo}
\newacronym{wds}{WDS}{Washington Double Star catalog}
\newcommand{\edd}{}  % edited first iteration
\newcommand{\edl}{}  % edited from language editor. First and second iteration

\usepackage[normalem]{ulem}
\usepackage{color}
\graphicspath{ {./images/} }
\begin{document} 

   \title{Radial-velocity jitter of stars as a function of observational timescale and stellar age\thanks{Based on data obtained from the ESO Science Archive Facility under request number sbrems392771.}$^{,\,}$\thanks{\edd{Table E.1 is only available at the CDS via anonymous ftp to \url{cdsarc.u-strasbg.fr} (\url{130.79.128.5}) or via \url{http://cdsarc.u-strasbg.fr/viz-bin/cat/J/A+A/632/A37}}}}
   %\ of young stars}

  % \subtitle{Dependence of Stellar Jitter on Age and Timescale}

   \author{Stefan S. Brems\inst{1} \and
          Martin K\"urster\inst{2} \and
          Trifon Trifonov\inst{2} \and
          Sabine Reffert\inst{1} \and
          Andreas Quirrenbach\inst{1}
          }
   \offprints{S. S. Brems, \email{sbrems@lsw.uni-heidelberg.de}}
   \institute{Zentrum f\"ur Astronomie der Universit\"at Heidelberg, Landessternwarte, K\"onigstuhl 12, 69117 Heidelberg, Germany
             %\email{sbrems@lsw.uni-heidelberg.de}
         \and
             Max Planck Institut f\"ur Astronomie, K\"onigstuhl 17, 69117 Heidelberg, Germany
             %\email{c.ptolemy@hipparch.uheaven.space}
             %\thanks{The university of heaven temporarily does not accept e-mails}
             }

   \date{Received 22 March, 2019 / accepted 08 October, 2019}

% 5 {} token are mandatory

  % context heading (optional)
  % {} leave it empty if necessary  
\abstract{Stars show various amounts of \gls{rv} jitter due to varying stellar activity levels. The typical amount of \gls{rv} jitter as a function of stellar age and observational timescale has not yet been systematically quantified, although it is often larger than the instrumental precision of modern high-resolution spectrographs used for Doppler planet detection and characterization.}
%For modern \gls{rv} instruments such as HARPS, CARMENES or ESPRESSO, the limiting factor for the identification of new planets is often not instrumental precision but rather stellar activity. Thus one either has to understand and remove this activity, or try to estimate the stellar noise contribution beforehand so it can be minimized by adjusting the observational strategy.%too active stars can be avoided.  The latter approach is the basic idea of this paper.}
  % aims heading (mandatory)
   {We aim to empirically determine the intrinsic stellar \gls{rv} variation \edd{for mostly G and K dwarf stars} on different timescales and for different stellar ages independently of stellar models. \edd{We also focus on young stars ($\lesssim 30$ Myr), where the \gls{rv} variation is known to be large.}}
  % methods heading (mandatory)
   {We use archival FEROS and HARPS \gls{rv} data of stars which were observed at least 30 times spread over at least two years.  We then apply the \gls{pv} technique to these data sets to identify the periods and amplitudes of underlying, quasiperiodic signals. We show that the \gls{pv} is a powerful tool to identify quasiperiodic signals in highly irregularly sampled data sets.}
  % results heading (mandatory)
   {We derive activity-lag functions for \edd{20 putative single} stars, where lag is the timescale on which the stellar jitter is measured. Since the ages of all stars are known, we also use this to formulate an activity--age--lag relation which can be used to predict the expected RV jitter of a star given its age and the timescale to be probed. The maximum \gls{rv} jitter \edd{on timescales of decades} decreases from \edd{over 500 m/s for 5 Myr}-old stars \edd{to 2.3 m/s for stars with ages of around 5 Gyr}. The decrease in \gls{rv} jitter when considering a timescale of only 1~d \edd{instead of 1~yr} is smaller by roughly a factor of \edd{4} for stars with an age of about \edd{5 Myr}, and a factor of \edd{1.5} for stars with an age of \edd{5}~Gyr. The rate at which the \gls{rv} jitter increases with lag strongly depends on stellar age and reaches 99\% of the maximum \gls{rv} jitter over a timescale of a few days for  stars that are a few million years old, up to \edd{presumably} decades \edd{or longer} for stars with an age of a few gigayears.}
  % conclusions heading (optional), leave it empty if necessary 
   {}

   \maketitle
%
%-------------------------------------------------------------------
\section{Introduction}\label{sec:introduction}
\glsresetall
Of the almost 4000 exoplanets known today, more than 3700 were discovered via the
\gls{rv}  or transit methods\footnote{\label{nasa_exoarchive}\url{https://exoplanetarchive.ipac.caltech.edu}}. According to the database, only three of them (V830 Tau b, \citealt{Donati2016}; K2-33 b, \citealt{David2016}, and TAP 26 b, \citealt{Yu2017})
%CI Tau b \edit, CVSO 30 b \edit, 1SWASP J1407 b\edit and CoRoT-20 b\edit
 are  younger than 100 Myr. The main reason for this is the strong stellar activity of young stars, which makes it hard to find the subtle planetary signal in the large stellar variations. This is unfortunate for two reasons: First, planet formation takes place in young systems and at least gas giants need to form before the disk has dissipated after less than a few tens of millions of years \citep[e.g.,][]{Ercolano2017}. Second, planets at large orbital distances ($\gtrsim 50$ AU) are almost exclusively detected via \gls{di}, which is best applicable to young systems where the planets are still hot from their formation. Thus, in order to discover all planets in a system, one either needs to image old stars -- which seems currently impossible given the already low detection rate around young stars \edd{probed by large \gls{di} surveys \citep[e.g.,][]{Desidera2015,Lannier2016,Tamura2016,Stone2018}} -- or try to minimize the impact of the stellar activity of young stars. A lot has been done to understand and characterize stellar activity \citep[e.g.,][]{Dumusque2018}.
\edd{\citet{Lindegren2003} further estimate the effects of stellar activity such as oscillation, granulation, meridional flow, long-term magnetic cycle, surface magnetic activity and rotation, gravitational redshift and many more on the \gls{rv} measurement. \citet{Meunier2019} then try to model the effect of this kind of activity signal on \gls{rv} data of mature stars. With our data probing activity timescales of days to years, we are mainly probing the combined effect of stellar rotation, reconfiguration of active regions, and  long-term magnetic cycles.} 
 Still, large uncertainties remain in the prediction and interpretation of any \gls{rv} signal, in particular for young pre-main sequence stars. But since this is what we measure, knowledge about the typical \gls{rv} variability is important, for example for developing and testing \gls{rv} activity models or planning \gls{rv} surveys. In this paper we therefore derive a model-free analytic relation between stellar jitter, stellar age, and lag, where lag denotes the timescale on which the jitter is measured.% We can show that the stellar noise increases by a factor of about two when probing timescales of months or years, compared to timescales of a few days only. Additionally we quantify the known, very strong dependence of stellar jitter on age. For 10 Myr old stars the jitter is larger by one order of magnitude than for 100 Myr old stars, and two orders of magnitudes larger when compared to 10 Gyr old stars.

In order to derive this relation, we systematically analyze precise Doppler measurements from FEROS \citep{Kaufer1999} and HARPS \citep{Mayor2003} \edd{of mainly G and K dwarfs with ages between 3 Myr and 7 Gyr} using the \gls{pv} technique \edd{\citep{Donahue1997a,Donahue1997b,Kuerster2004}}.

%Still \citet{WeisePhd2010} and \citet{MohlerFischer2013} were not able to identify any planetary signal in their data sets of 121 young ($\lesssim 100$ Myr, many $\le 10$ Myr), late (F6--M2) stars. In this paper we show that the stellar activity contribution of young stars can be reduced by a) roughly a factor of two when searching for planets with very short orbital periods of a few days maximum and b) about an order of magnitude by targeting stars older than about 10 Myr, but still younger than 100 Myr. To do this we use the \gls{pv} method. We use archival data from HARPS \citep{Mayor2003} and FEROS \citep{Kaufer1999}. Apart from smaller activity amplitudes, a second advantage of probing shorter periods is that the \gls{rv} signal will increase towards shorter periods. This will then minimize the planet's minimum mass needed for a detection via \gls{rv}. The first survey known to the authors making use of this is the RV\,SPY survey led by Olga Zakhozhay (Zakhozhay et al., in prep.).
\smallskip
This paper is organized as follows: In Sect. \ref{sec:target_sample} we introduce the input target list as well as target selection and data cleaning. Section \ref{sec:method} explains the \gls{pv} technique, the activity modeling for individual stars, and the uncertainty estimation. Section \ref{sec:results} presents two analytic activity--age--lag functions to the pooled data of all stars. In Sect. \ref{sec:discussion} we discuss the strengths, weaknesses, and limits of this analysis and give example values of the empirical activity--age--lag function. Section~\ref{sec:conclusions} then concludes on the main findings of this paper.
%--------------------------------------------------------------------

\section{Target sample}\label{sec:target_sample}

\begin{table*}
        \caption{Properties of the 27 stars that qualified for further analysis. \edd{Column "Bin. Sep." lists the projected separation of the binary companion, if present, to the host star as listed in the \gls{wds}. The acronym SB refers to a spectroscopic binary identified in our data.}}
        \begin{tabular}{lcccccccccc}
        \hline \hline
        Main ID & RA & DEC & SpT & Age & Instrument & Catalog & Age ref.&\edd{Bin. Sep.} \\
 &  &  &  & $\mathrm{[Myr]}$ &  &  & &\edd{[AU]}\\
\hline
\object{V2129 Oph} & 16:27:40.286 & --24:22:04.030 & K7 & $3\pm2$ & FEROS & 1, 2 & 4&78 \\%EM* SR 9 is listed as 5\pm3 Myr in Weise+10
\object{HD 140637} & 15:45:47.600 & --30:20:56.000 & K2V & $5\pm2$ & FEROS & 2 & 5 &27\\%KW Lup
\object{CD-37 13029} & 19:02:02.000 & --37:07:44.000 & G5 & $5\pm2$ & FEROS & 2 & 5 \\%V702 CrA
\object{TYC 8654--1115--1} & 12:39:38.000 & --57:31:41.000 & G9V & $5\pm3$ & FEROS & 2 & 4 \\%CD--56 4581
\object{HBC 603} & 15:51:47.000 & --35:56:43.000 & M0 & $5\pm3$ & FEROS & 2 & 4 &280\\%SZ 77
\object{HD 81544} & 09:23:35.000 & --61:11:36.000 & K1V & $8\pm3$ & FEROS & 2 & 4 &591\\%V479 Car
\object{TYC 5891--69--1} & 04:32:43.509 & --15:20:11.268 & G4V & $10\pm5$ & FEROS & 2 & 4 \\%2MASS J04324350--1520114
\object{CD--78 24} & 00:42:20.300 & --77:47:40.000 & K3V & $15\pm10$ & FEROS & 2 & 4 \\
\object{1RXS J043451.0--354715} & 04:34:50.800 & --35:47:21.000 & K1V & $20\pm15$ & FEROS & 2 & 4 \\ %CD--36 1785
\object{1RXS J033149.8--633155} & 03:31:48.900 & --63:31:54.000 & K0V & $25\pm15$ & FEROS & 2 & 4 \\ % TYC 8870--372--1
\object{TYC 7697--2254--1} & 09:47:19.900 & --40:03:10.000 & K0V & $25\pm10$ & FEROS & 2 & 4 \\%CD--39 5833  
\object{CD--37 1123} & 03:00:46.900 & --37:08:02.000 & G9V & $30\pm15$ & FEROS & 2 & 4 \\
\object{CD--84 80} & 07:30:59.500 & --84:19:28.000 & G9V & $30\pm15$ & FEROS & 2 & 4 \\
\object{HD 51797} & 06:56:23.500 & --46:46:55.000 & K0V & $30\pm15$ & FEROS & 2 & 4 \\
\object{TYC 9034--968--1} & 15:33:27.500 & --66:51:25.000 & K2V & $30\pm15$ & FEROS & 2 & 4 \\%1RXS J153328.4-665130
\object{HD 30495} & 04:47:36.210 & --16:56:05.520 & G1.5 V & $45\pm10$ & HARPS & 1, 3 & 6 \\ %58 Eri
\object{HD 25457} & 04:02:36.660 & --00:16:05.920 & F7 V & $50\pm15$ & Both & 1, 2, 3 & 4 \\
\object{1RXS J223929.1--520525} & 22:39:30.300 & --52:05:17.000 & K0V & $60\pm15$ & FEROS & 2 & 4 \\ %CD--52 10232
\object{HD 96064} & 11:04:41.580 & --04:13:15.010 & G5 & $90\pm10$ & FEROS & 2 & 4 &123\\
\object{HD 51062} & 06:53:47.400 & --43:06:51.000 & G5V & $200\pm50$ & FEROS & 2 & 4 &SB\\
\object{HD 202628} & 21:18:27.269 & --43:20:04.750 & G5 V & $604\pm445$ & HARPS & 1, 3 & 7 \\
\object{HD 191849} & 20:13:52.750 & --45:09:49.080 & M0 V & $850\pm400$ & HARPS & 1, 3 & 8 \\
\object{HD 199260} & 20:56:47.331 & --26:17:46.960 & F6 V & $3460$ & HARPS & 1, 3 & 9 \\
\object{HD 1581} & 00:20:01.910 & --64:52:39.440 & F9 V & 3950 & HARPS & 1 & 8 \\ %$\zeta$ Tuc
\object{HD 45184} & 06:24:43.880 & --28:46:48.420 & G2 V & 4420 & HARPS & 3 & 10 \\
\object{HD 154577} & 17:10:10.270 & --60:43:48.740 & K0 V & $4800$& HARPS & 3 & 10 \\
\object{HD 43834} & 06:10:14.200 & --74:45:09.100 & G5 V & $7244\pm3226$ & HARPS & 1 & 11 &32\\ % $\alpha$ Men

        \hline
        \end{tabular}
        \label{tab:stellarproperties}
        \tablebib{(1): I\,SPY Launhardt et al. (in prep.); (2): \citet{MohlerFischer2013} and \citet{Weise2010}; (3): RV\,SPY Zakhozhay et al. (in prep.).; (4) \citet{Weise2010}; (5) \citet{WeisePhd2010}; (6) \citet{Maldonado2010}; (7) \citet{TucciMaia2016}; (8) \citet{Vican2012}; (9) \citet{Ibukiyama2002}; (10) \citet{Chen2014}; (11) \citet{Lachaume1999};}
        \end{table*}

%textwidth: \printinunitsof{cm}\prntlen{\textwidth}
%linewidth: \printinunitsof{cm}\prntlen{\linewidth}
\begin{figure}[ht]
        \includegraphics[width=\linewidth]{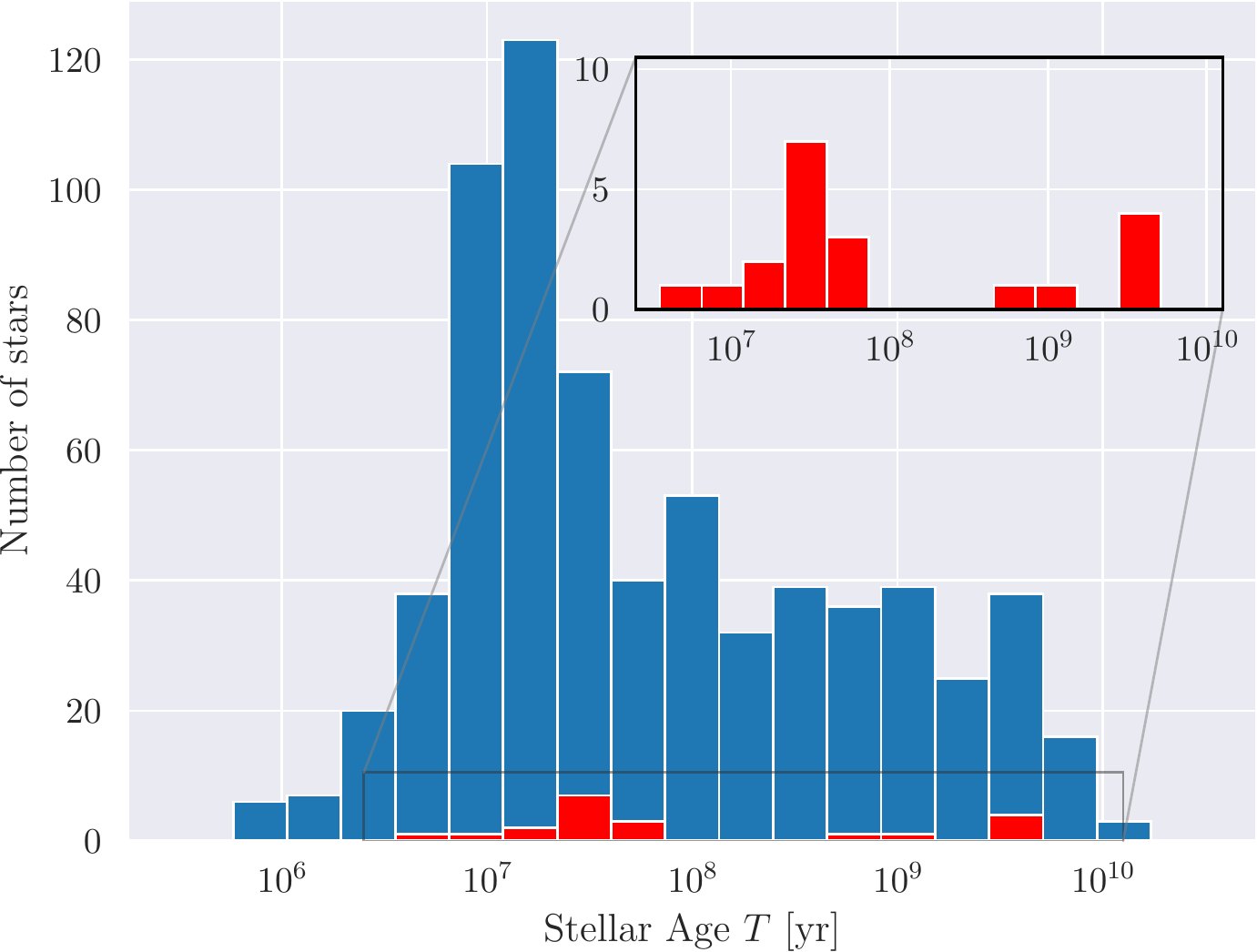}
        \caption{Age distribution of the stars in the input catalog (blue) and the remaining \edd{20 single} stars (red) after applying the selection criteria described in the text. The remaining sample has a \edd{wide age distribution, slightly skewed towards younger stars}. Young stars ($\leq 100$ Myr) are especially important for this analysis and are in general rarely observed in \gls{rv} surveys.}

        \label{fig:agedist_all}
\end{figure}

%One of our goals is to optimize \gls{rv} surveys for directly imaged stars and also get an age-activity dependence. Since stars used for \gls{rv} surveys are typically very old when compared to directly imaged stars, we specifically tried to find surveys focusing on young stars. This has the advantage, that we have stars with known ages, and also get a wide age spread, since old stars will be present in almost every \gls{rv}-survey.
%since o which are typically very young, our main criterion is the known age and youth of the stars. 
Our goal is to characterize \gls{rv} jitter as a function of stellar age and probed timescale. Therefore, we need to put together a sample of young and old stars with known ages that have been part of \gls{rv} monitoring programs.

We assemble our input target list from three different surveys which all focus on young stars in the southern hemisphere: \edl{First, the ongoing NaCo-ISPY\footnote{\url{http://www.mpia.de/NACO_ESPRI_GTO}} (Imaging Survey for Planets around Young stars) survey (Launhardt et al., in prep.; 443 targets). Second the recently started RV\,SPY (Radial-velocity Survey for Planets around Young stars) survey (Zakhozhay et al., in prep.; 180 targets). And third the \gls{rv} survey to find planets around young stars from \citet{WeisePhd2010} and \citeauthor{MohlerFischer2013} (\citeyear{MohlerFischer2013}; 214 targets).} \edd{NaCo-ISPY  is a 120-night guaranteed time observations (GTO) $L'$-band direct imaging survey using the NaCo instrument \citep{Lenzen2003,Rousset2003} at the VLT. Its target list contains mostly two subcategories: First, young, nearby stars with debris disks showing significant infrared excess, and second, stars with protoplanetary disks. \edl{RVSPY is a \gls{rv} survey currently} spanning 40 nights that is complementary to NaCo-ISPY, as it searches for close-in planets around debris disk stars older than about 10~Myr. Its target list has an intentionally large overlap with the NaCo-ISPY target list, but is extended especially for older, less active stars. Since the second and third surveys use the FEROS instrument and young stars are mostly avoided in \gls{rv} surveys, the young stars ($\leq 45$ Myr) are uniquely observed with FEROS; see Table \ref{tab:stellarproperties}. However, since their intrinsic \gls{rv} variation is at least one order of magnitude larger than the instrumental precision, we do not expect any significant bias in our results from this selection effect.} Since there are \edd{overlaps} in the catalogs, we end up with 699 targets. Figure \ref{fig:agedist_all} shows the age distribution of those stars. One can see that most stars in the input catalog are younger than 100 Myr. Almost all of the older stars come from the RV\,SPY survey, which also included older stars to avoid the issues of young and active stars as \gls{rv} targets.

We searched the archives for public data from the ESO-instruments HARPS \citep{Mayor2003} and FEROS \citep{Kaufer1999} for all 699 stars in our input catalog.
%765 targets (ignoring duplicates and stars without determined ages). 
The \gls{rv}s were derived using the CERES pipeline \citep{Brahm2017} for FEROS and the SERVAL pipeline \citep{Zechmeister2018} for HARPS data. We removed bad observations with formal errors \edd{as returned by the pipelines} above 20 m/s and 50 m/s for HARPS and FEROS respectively, and via iterative $10\,\sigma$-clipping on all data of a star simultaneously. In order to qualify for our final analysis, the remaining data need to be sufficiently evenly distributed for each star and instrument. We ensured this by requiring a minimum of 30 observations, at least a two-year baseline, no gap larger than 50\%, and a maximum of two gaps larger than 20\% of the baseline. Since HARPS underwent a major intervention, including a fiber change in June 2015 \citep{LoCurto2015}, these criteria needed to be fulfilled for one of the data sets before or after the intervention.

Finally stars with known companions (stellar or substellar) listed in the Washington Double Star catalog \citep{Mason2001}, the Spectroscopic Binary Catalog \citep[9th edition][]{Pourbaix2004}, or the NASA Exoplanet Archive$^1$%\cref{nasa_exoarchive}
 were also removed. This is done so we can make the assumption that we are left only with stellar noise.  \edd{As a consistency check with increased statistics, we additionally analyzed seven of the removed wide binary systems ($\ge 27$ AU projected separation). They are not used to obtain the main results given in Table \ref{tab:mainresults} and are marked with a binary flag in Tables \ref{tab:stellarproperties} and \ref{tab:mainresults}.} 

After this selection we were left with 27 stars \edd{(including the 7 binaries)}: 9 with sufficient HARPS data and 19 with sufficient FEROS data, where %HD~183414 and 
HD~25457 \edd{(single)} had good data from both instruments. Table \ref{tab:mainresults} lists their basic properties and  Fig. \ref{fig:all_rvdata} shows the \gls{rv} data for all 27 stars. \edd{The individual measurements, including the barycentric Julian date (BJD), the \gls{rv} signal, and the \gls{rv} error, are given in Table E.1. These data are only available electronically at the CDS.}
%Each instrument and star then had to fulfill the criteria  As this final selection based on spacing is hard to quantify, it was performed manually. We were left with \edit stars, having a median of \edit observations each. For \edit stars we found enough data for both instruments. For HARPS we then used the reduced data obtained from the official HARPS \gls{drs}, whereas we reduced all the FEROS data using the CERES pipeline \citep{Brahm2017}. Only for HD~25457 we used SERVAL \citep{Zechmeister2018} for the HARPS data, as the HARPS \gls{drs} produced unphysical jumps in the RV data, which are most likely due to different input parameters the different users gave for the different observing programs.
\section{Method}\label{sec:method}
\begin{figure*}[h]
  \includegraphics[width=\textwidth]{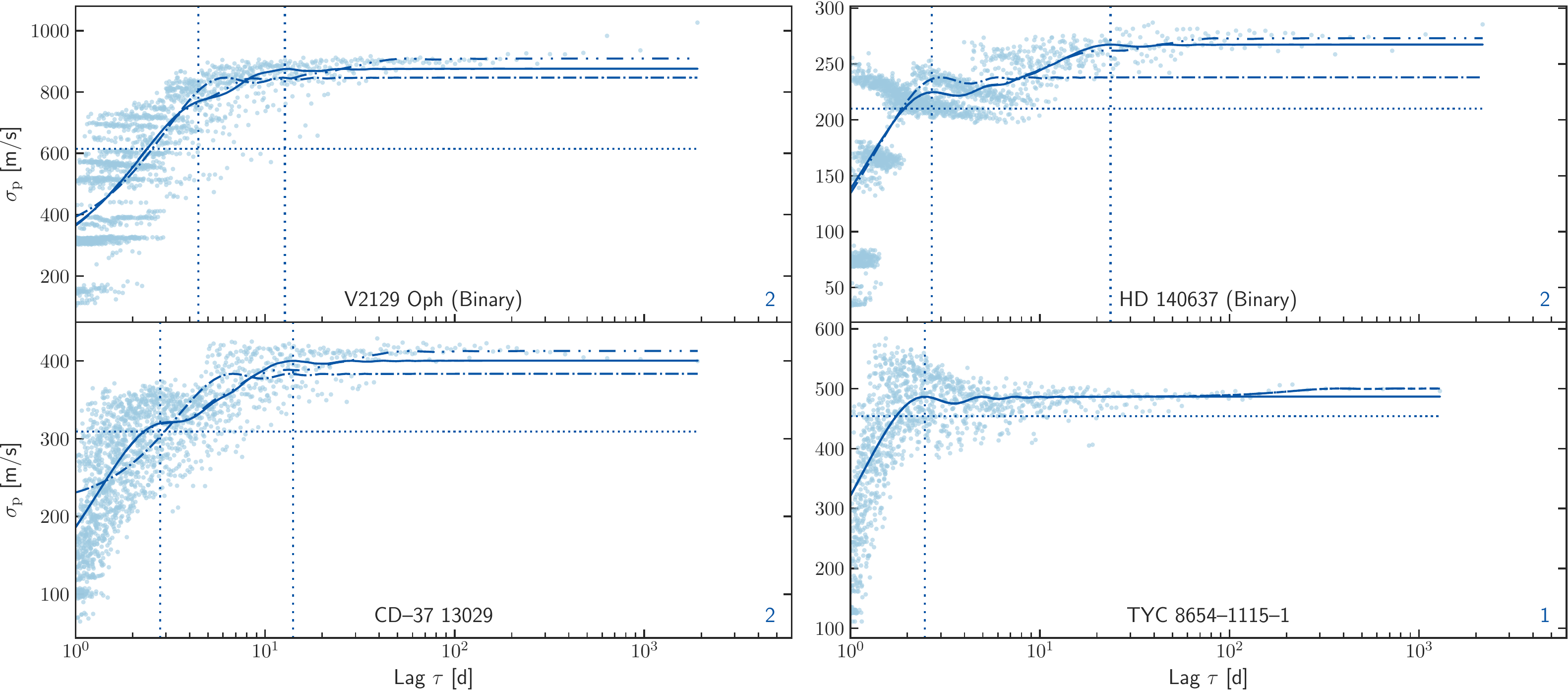}
  \caption{Pooled results for the first 6 stars. Find plots for the remaining 21 stars in Fig. \ref{fig:all_results_appendix}. Blue symbols denote FEROS data and green symbols HARPS data. The lines show the best fits for the different numbers of sinusoidal signals fitted: Dotted: zero signal (constant), dash-dotted: one signal, dashed: two signals, dash-dot-dotted: three signals. The solid line replaces the line which qualified as best fit using the F-test described in Sect. \ref{sec:modeling}. \edd{The dotted vertical lines represent the periods of the identified signals and the number in the bottom right corner the number of identified signals.} The model parameters are shown in Table \ref{tab:mainresults}. We note that $\sigma_\text{p}$ and not $\sigma_\text{p}^2$ is shown.}
  \label{fig:all_results_text}
\end{figure*}

\subsection{Models}
In order to determine the typical \gls{rv} scatter over different observing timescales (lags), several methods were applied.% For the record we list all methods we tried:

\edl{First, the variogram or structure function \citep{Hughes1992} using different estimators as described in \citet{Rousseeuw1993} and \citet{Eyer1999} was tried: The idea here is to create all possible differences between the measured \gls{rv} data points. Those points are then sorted by the time differences (lags) and compared to theoretical predictions of a sinusoidal signal. Second, self-created, automated block-finding algorithms with random selection of single observation in clustered observations: The algorithm identifies clustered observations on different timescales. For clusters shorter than an arbitrarily chosen fraction of the lag probed, one random observation will be picked, while longer blocks will be split into sub-blocks of the according size. For each block, the variance is determined and used as the typical variance for that timescale. Third, consecutive binning was tried, where the observations are binned on the timescale chosen. This is similar to the second method, but there is no upper limit on the block size and one takes the mean instead of a random representative of that bin. This is done for all lags to be probed. And finally the \gls{pv} was used.}
%But out of these methods only the \gls{pv} was capable of identifying the expected increase of \gls{rv} scatter with lag. To make sure that this increase is real, we created 100 random  data sets with zero to three sinusoidal signals buried in them and looked how the \gls{pv} algorithm recovers the known signals. The algorithm recovers them well as can be seen in Fig. \ref{fig:randomsampling}.

Since only the latter method turned out to be robust enough to identify signals in sparse and irregularly sampled data, we used it for our analysis. The remainder of this section is dedicated to describing the method in more detail.
\subsection{Pooled variance}\label{sec:pv}
We use the  \gls{pv} or \gls{pvd} method, which was first introduced for the analysis of time series of astronomical data by \citet{Dobson1990} to analyze the CaII emission strength of active late-type stars. \citet{Dobson1990}, \citet{Donahue1995}, \citet{Donahue1996}, \citet{Donahue1997a, Donahue1997b}, and \citet{Kuerster2004} demonstrated the capability of this technique to detect timescales pertaining to stellar activity such as \edl{the stellar rotation period, the typical timescale of active region reconfiguration or the stellar activity cycle.} The \gls{pv} is a combined variance estimate from $k$ different sets of measurements $y_{i,j}, \,i = 1,\dots, N_j, \,j=1,\dots,\,k$ each of which has a different mean $\overline{y_j}$. If it can be assumed that all the individual sets of measurements have the same variance (despite the different mean), then the \gls{pv} $\sigma_\text{p}$ is defined as
\begin{align}\label{eq:pvdefinition}\begin{split}
\sigma^2_{\textrm{p}, k} :&= \frac{(N_1-1)\sigma_1^2+\dots+(N_k-1)\sigma_k^2}{(N_1-1)+\dots+(N_k-1)}\\
&=\frac{\sum_{i=1}^{N_1}(y_{i,1}-\overline{y_1})^2+\dots+\sum_{i=1}^{N_k}(y_{i,k}-\overline{y_k})^2}{N_1+\dots+N_k-k},
\end{split}\end{align}
with $\sigma_{\text{p}, j}$ being the variance of the $j$th data set, that is, the \gls{pv} is the weighted mean of the variances of the individual data sets.

In this paper we make the assumption of (on average) equal \gls{rv} variances in spite of different mean \gls{rv} for data sets that were obtained within time intervals of equal length. In particular we assume this to be true for the data taken by HARPS before and after the intervention in June 2015 \citep{LoCurto2015}\edd{, even though the formal errors increase from a mean of 0.9 m/s to 1.3 m/s for our data. This decreased precision after the intervention seems to affect all HARPS data (Trifonov in prep.), but since we probe jitter values typically much larger than these formal errors, the assumption still holds. The only significant effect is the offset of several meters per second in the absolute \gls{rv} values. To account for this, we split the data into pre- and post-data sets.} \edd{Furthermore, we assume that the \gls{rv} variance on a given timescale does not change over time (e.g., due to activity cycles).}  For these data sets the \gls{pv} can be calculated and is used as an estimate of the true variance for this timescale. As long as the lag probed is shorter than the observational baseline, the \gls{pv} is more precise than the variance of a single data set due to the larger number of measurements as several data sets are combined. If the length of the time intervals differs we expect different variances in general.
% As, if we now split the measurements series into smaller and smaller blocks (increasing $k$), the number of observations per block ($N_j,\,j=1,\dots,k$) will go down. As a consequence, the mean will change in a way that the variance will also systematically decrease, compare e.g. \citet{Mawet2014}. To correct for this effect, we add another factor $1+\frac{1}{N_j-1},\,j=1,\dots,k$ to each of the blocks:
% \begin{equation}
% \sigma^2_{\textrm{p}, k,  \text{corr}} :&= \frac{(N_1-1)\sigma_1^2\cdot(1+\frac{1}{N_1-1})+\dots+(N_k-1)\,\sigma_k^2\cdot(1+\frac{1}{N_k-1})}{(N_1-1)+\dots+(N_k-1)}
% \end{equation}
% We solely use this corrected version throughout the paper for determining the \gls{pv}. Also we do not use the simple variance $\sigma_i^2$, but we weight each point by the inverse of its measurement uncertainty given by the reduction pipeline.

\subsection{Block sizes} \label{sec:blocksizes}
As the integer $k$ to split our observations in smaller blocks is arbitrary, we run $k = 1,\dots, B$, where $B$ is the number of days covered by the observations. In other words, we split our data into blocks, starting from one block having the full baseline length, to $B$ blocks with a length of one day each. We define the lag $\tau := B/k$, which corresponds to the length of each bin in days for a given $k$. For each value of $k$, or equivalent lag $\tau$, we then obtain a different variance $\sigma_{\textrm{p}, k}^2$ as defined in Eq. \ref{eq:pvdefinition}. We note that due to the HARPS fiber change in June 2015 \citep{LoCurto2015}, we removed HARPS data taken during this procedure (2457173.0 -- 2457177.0 JD) and did not allow blocks to combine data taken before and after the intervention. We therefore had two data sets with shortened baselines. After applying the \gls{pv}, we then treat the two datasets as one again.\\
%\textit{Note:} Note every block has to return a value, as especially for small blocks a minimum number of two values is required to calculate the standard deviation. If there was e.g. only one measurement, this one measurement will not contribute to this $\sigma_{\textrm{p}, k}^2$ then.
%We assign each $\sigma_{\textrm{p}, k}^2$ a weight given by the square-root of the product of the number of valid blocks and the total number of used measurements. 

%\FloatBarrier
It is difficult to assign absolute error bars to each pooled data point, as the formal \gls{rv} precision (a few m/s) is typically much smaller than the \gls{pv} (a few hundred m/s). We therefore decided to assign a weight $w_k$ to each measurement $\sigma_{\textrm{p}, k}$. In contrast to errors, weights only have a relative meaning and thus do not need to be calibrated absolutely. Since more points yield in general a more significant result , we use the square root of the number of individual points minus the number of filled boxes that were used to compute this point. We subtract the number of filled boxes because each box takes one degree of freedom (the mean) :
\begin{equation}\label{eq:weights}
        \text{w}_k := \sqrt{\frac{n_{k}-\tilde{k}}{2}}\quad , 
\end{equation}
where $n_k := N_1+\dots+N_{\tilde{k}}$ is the number of individual data points contributing, and $\tilde{k}$ is the number of blocks where the variance could be calculated, that is, the number of boxes with at least two measurements. This formula, and especially the subtraction of $\tilde{k},$ are further motivated by the relative uncertainty of the variance estimator for $N$ points, which is $\sqrt{\frac{2}{N-1}}$ \citep[e.g.,][p. 22]{Squires2001}. Thus, under the assumption of $k$ independent blocks with $N/k$ measurements in each block, the uncertainty is
\begin{equation}
\sqrt{\frac{2}{N/k-1}}\cdot\frac{1}{\sqrt{k}} = \sqrt{\frac{2}{N-k}} \equiv w_k^{-1}
\end{equation}
for $k$ independent blocks with $N/k$ measurements in each block \citep[compare also][]{Brown2007}.
%\begin{equation}\textrm{w}_k = \sum_{i=0}^k\sqrt{n_{i, k}}\quad ,\end{equation}
%where $n_{i, k}$ is the number of measurements of the $i$th block.
% But it shows that assigning weights by using bootstrapping or the number of measurements used to generate the point does not change the results significantly, no weights were used.

\subsection{Activity modeling} \label{sec:modeling}
Now that we know the variance on different timescales, we want to find the periods $P$ and amplitudes $K$ of the underlying modulations. Under the assumption of an underlying, infinitely sampled sinusoidal signal $y(t) = K \sin \left( \frac{2\pi}{P}t -\delta \right)$ with semi-amplitude $K$, period $P,$ and phase $\delta$, the analytic \gls{pv} for timescale $\tau$ is given in equation \ref{eq:pvsine}, which is independent of $\delta$.

Since there might be no, one, or multiple periodic signals of this kind, we fit four different curves to each \gls{pv} data point of a star:
\begin{equation}\label{eq:multipv}
  \sigma_\text{p}^2(\tau) = A^2 + \sum_{i=1}^m K_i^2\left[ \frac12 + \frac{\cos(2\pi\tau/P_i)-1}{(2\pi\tau/P_i)^2} \right]\quad,
\end{equation}
where $m\in [0, 1, 2, 3]$ is the number of sinusoidal signals and $\tau$ is the timescale or lag probed. In other words, only white noise or white noise plus up to three underlying sinusoidal signals are fitted using $\chi^2$-minimization, where the formal squared errors are given by the reciprocal of the weight from Eq. \ref{eq:weights}. We note that the models have $2m+1$ free parameters. In order to decide how many signals are significant, we used an F-test \citep{Rawlings1998}, which is often used in nested models. We highlight the fact that the number of independent measurements required in this test is the number of observations and not the number of points in the \gls{pvd}. This approach has one value $\alpha$ to be chosen arbitrarily.
%free parameter 
This $\alpha$ acts as a threshold \edd{in rejecting the null hypothesis, which is that the model with fewer parameters describes the data as well as the model with more parameters.} %on the significance of the additional parameter and 
 It is typically chosen to be around $\alpha\approx 0.01-0.05$\edd{, where a larger value favors the model with fewer parameters.} %To have a rather conservative approach, 
We select $\alpha = 0.05$. % since we rather want to wrongly identify a signal than miss an increase of jitter.
%Finally, using the weights described above, we fit a powerlaw \begin{equation}
%\gamma(\Drv) = a\cdot \Drv^\alpha\end{equation}
% where $\gamma$ is the standard deviation to be expected for points spanning a baseline of \Drv\ and $a$ and $\alpha$ are free parameters. While $a$ represents the typical scatter of data points at shortest temporal timescales of a few days due to stellar activity, $\alpha$ represents the growth of this scatter due to more and more factors coming into play. The latter is exactly the value we are looking for.

\subsection{Uncertainty estimation}\label{sec:error}

Finally we want to assign confidence intervals to each of the $2m+1$ parameters we found for each star to best describe its activity.
We tried the following methods:

\edl{First, we tried to apply \gls{mcmc} on the \gls{rv} data as well as on the pooled data. Then we tried a \gls{mc}-like method on the pooled data: we removed the modeled signal from the pooled data, binned seven neighboring data points, for example, and determined their mean and variance. This mean and variance were then used to create seven new, random data points, assuming a Gaussian distribution. We then re-added the modeled signal and re-performed the fitting routine. And finally we tried bootstrap resampling on the \gls{rv} data as well as on the pooled data.}

Quantifying these methods not only by eye but also using artificially generated data where we know the true periods and amplitudes of the underlying signals, we found the last method to give the most realistic results. By "realistic", we mean that the true value lies within the error bars of the estimated values and that the sampled values spread roughly equally around the original estimate. The \glsentryshort{mcmc} methods underestimated the errors, whereas \edl{in the \gls{mc}-like method} one often has the problem of non-Gaussian distributed residuals, which will then lead to skewed and shifted results if those are approximated and redistributed by Gaussians. Bootstrapping the original data leads to issues when bins often contain identical data, returning a variance of zero, seemingly skewing the results to smaller absolute variances.

%More specifically: 
For the last method, we pooled the data using Eq. (\ref{eq:pvdefinition}) as described in Sects. \ref{sec:pv} and \ref{sec:blocksizes}, resulting in $k$ pairs of lag $\tau$ and variance $\sigma_\text{p}^2$.  Subsequently, we determined the model capable of best describing the data using the F-test; see Sect. \ref{sec:modeling}. We randomly redrew $k$ pairs of lag $\tau$ and variance $\sigma_\text{p}^2$ with placing back, meaning the same pair can be drawn mutliple times. This is the method known as \textit{bootstrapping}. Fixing the model to the one found for the original data, we fit this model to the new data. Repeating the last two steps 5000 times yields an estimate of the robustness of the model parameters. The original fit is used as the best fit and the standard deviations left and right of that \edd{are used as} the asymmetric error bars. An example of this distribution for HD~45184 is shown in Fig. \ref{fig:bootstrcornerplot} (bottom left), where the lines indicate the best fits and confidence intervals. However, bootstrapping assumes the underlying data points to be mutually independent and well sampled. Since the statistical independence of the \glspl{pv} is not fulfilled here and additional systematic errors might be present, the errors should be seen as lower limits to the real uncertainties.%: One without a signal, but only Gaussian and one with so called \textit{red noise}: Two sinusoidal signals with periods of 7 d and 365 d and semi-amplitudes of 0.4 and 0.6 times that of the added Gaussian noise. Given the last approach, we could recover the signal with \edit.

 %In order to do this, we use an \gls{mc} like approach: First we removed the signal analytically described in Eq. (\ref{eq:multipv}) from the pooled data. Assuming that there is only Gaussian noise left, we binned 10 neighboring data points to determine the mean and standard deviation of $\sigma^2(\tau)$ each bin using the same weights as always. Without changing the lag $\tau$ of each point, we randomly drew 10 data points from the resulting normal distribution and assigned them to the data points. Afterwards we added the analytical signal we removed in the first place. We repeated this process 5000 times and then determined the 68\% confidence intervals around the original best fit as errors. Please note that we can only give lower limits to the real error, as cannot account for violations of the assumption, that each signal is densely sampled, e.g. each phase of it is covered. 

%--------------------------------------
\section{Results}\label{sec:results}%\glsresetall

{
\renewcommand{\arraystretch}{1.4} % Default value: 1
\begin{sidewaystable*}
        \begin{tabular}{lccccccccccccc}
        \hline\hline
        Main ID & \edd{Age} & Binary & N sines & Baseline & \#Obs & $A$ & $K_1$ & $P_1$ & $K_2$ & $P_2$ & $K_3$ & $P_3$ \\
 & \edd{Myr} & & & d &  & $\mathrm{m\,s^{-1}}$ & $\mathrm{m\,s^{-1}}$ & $\mathrm{d}$ & $\mathrm{m\,s^{-1}}$ & $\mathrm{d}$ & $\mathrm{m\,s^{-1}}$ & $\mathrm{d}$ \\
\hline
\object{V2129 Oph} & $3 $ & B & 2 & 1910 & 100 & $248.62^{23.29}_{45.49}$ & $937.34^{50.14}_{62.33}$ & $4.45^{0.40}_{0.42}$ & $729.75^{71.21}_{64.91}$ & $12.72^{6.21}_{1.73}$ & \dots & \dots \\
\object{HD 140637} & $5$ & B & 2 & 2167 & 63 & $0.00^{0.00}_{0.00}$ & $315.04^{1.13}_{1.10}$ & $2.69^{0.04}_{0.04}$ & $209.07^{3.09}_{3.02}$ & $23.59^{1.55}_{1.34}$ & \dots & \dots \\
\object{CD--37 13029} & $5$ &  & 2 & 1914 & 76 & $0.00^{0.00}_{0.00}$ & $434.00^{4.84}_{5.00}$ & $2.80^{0.05}_{0.05}$ & $363.52^{5.80}_{5.57}$ & $14.03^{1.01}_{0.83}$ & \dots & \dots \\
\object{TYC 8654--1115--1} & $5$ &  & 1 & 1290 & 70 & $0.00^{0.00}_{0.00}$ & $688.53^{1.60}_{1.62}$ & $2.47^{0.03}_{0.03}$ & \dots & \dots & \dots & \dots \\
\object{HBC 603} & $5$ & B & 1 & 774 & 51 & $535.81^{5.28}_{5.82}$ & $565.18^{11.45}_{10.87}$ & $9.67^{1.32}_{0.94}$ & \dots & \dots & \dots & \dots \\
\object{HD 81544} & $8$ & B & 1 & 1232 & 56 & $0.00^{40.75}_{0.00}$ & $234.48^{0.85}_{8.25}$ & $2.63^{0.17}_{0.04}$ & \dots & \dots & \dots & \dots \\
\object{TYC 5891--69--1} & $10$ &  & 3 & 1924 & 51 & $0.00^{0.00}_{0.00}$ & $248.41^{5.02}_{4.68}$ & $2.43^{0.06}_{0.05}$ & $251.86^{4.58}_{4.84}$ & $8.51^{0.60}_{0.46}$ & $247.06^{3.99}_{4.27}$ & $168.18^{13.36}_{13.32}$ \\
\object{CD--78 24} & $15$ &  & 1 & 1157 & 44 & $33.89^{42.48}_{31.67}$ & $285.13^{3.96}_{20.02}$ & $2.11^{0.09}_{0.06}$ & \dots & \dots & \dots & \dots \\
\object{1RXS J043451.0--354715} & $20$ &  & 1 & 1489 & 48 & $69.75^{1.63}_{1.66}$ & $246.37^{1.33}_{1.35}$ & $7.90^{0.18}_{0.17}$ & \dots & \dots & \dots & \dots \\
\object{1RXS J033149.8--633155} & $25$ &  & 1 & 1541 & 62 & $54.81^{0.94}_{1.01}$ & $141.54^{0.82}_{0.81}$ & $5.92^{0.12}_{0.12}$ & \dots & \dots & \dots & \dots \\
\object{TYC 7697--2254--1} & $25$ &  & 0 & 1458 & 49 & $107.62^{0.38}_{0.39}$ & \dots & \dots & \dots & \dots & \dots & \dots \\
\object{CD--37 1123} & $30$ &  & 1 & 1536 & 54 & $10.58^{0.17}_{0.19}$ & $18.11^{0.21}_{0.20}$ & $4.85^{0.18}_{0.17}$ & \dots & \dots & \dots & \dots \\
\object{CD--84 80} & $30$ &  & 1 & 690 & 35 & $0.00^{11.45}_{0.00}$ & $138.19^{1.76}_{1.88}$ & $5.42^{0.30}_{0.24}$ & \dots & \dots & \dots & \dots \\
\object{HD 51797} & $30$ &  & 1 & 682 & 48 & $0.00^{25.08}_{0.00}$ & $201.15^{1.23}_{3.42}$ & $4.04^{0.23}_{0.13}$ & \dots & \dots & \dots & \dots \\
\object{TYC 9034--968--1} & $30$ &  & 3 & 1269 & 92 & $27.83^{3.23}_{13.27}$ & $62.68^{5.12}_{2.53}$ & $3.55^{0.71}_{0.64}$ & $99.43^{2.70}_{5.03}$ & $10.46^{0.97}_{0.54}$ & $72.25^{2.40}_{2.65}$ & $171.35^{25.49}_{20.62}$ \\
\object{HD 30495} & $45$ &  & 2 & 1963 & 135 & $1.85^{0.09}_{0.11}$ & $9.08^{0.06}_{0.08}$ & $7.50^{0.19}_{0.21}$ & $15.78^{1.80}_{2.09}$ & $559.61^{117.15}_{115.34}$ & \dots & \dots \\
\object{HD 25457 (HARPS)} & $50$ &  & 1 & 4089 & 78 & $0.00^{-0.00}_{0.00}$ & $49.32^{0.08}_{0.07}$ & $4.54^{0.03}_{0.03}$ & \dots & \dots & \dots & \dots \\
\object{HD 25457 (FEROS)} & $50$ &  & 1 & 2273 & 45 & $17.54^{0.33}_{0.35}$ & $51.29^{0.27}_{0.25}$ & $5.08^{0.09}_{0.09}$ & \dots & \dots & \dots & \dots \\
\object{1RXS J223929.1--520525} & $60$ &  & 0 & 748 & 37 & $74.48^{0.36}_{0.37}$ & \dots & \dots & \dots & \dots & \dots & \dots \\
\object{HD 96064} & $90$ & B & 3 & 2946 & 124 & $2.05^{2.80}_{1.96}$ & $20.41^{0.32}_{0.42}$ & $2.36^{0.24}_{0.08}$ & $18.87^{0.49}_{0.94}$ & $6.35^{0.46}_{0.24}$ & $20.01^{0.70}_{0.74}$ & $240.25^{28.09}_{25.59}$ \\
\object{HD 51062} & $200$ & B & 1 & 1074 & 47 & $0.00^{348.53}_{0.00}$ & $8981.18^{121.44}_{124.71}$ & $15.01^{0.78}_{0.67}$ & \dots & \dots & \dots & \dots \\
\object{HD 202628} & $604$ &  & 1 & 1819 & 109 & $6.43^{0.04}_{0.04}$ & $9.07^{0.18}_{0.18}$ & $14.42^{1.07}_{0.81}$ & \dots & \dots & \dots & \dots \\
\object{HD 191849} & $850$ &  & 2 & 3229 & 39 & $1.87^{0.00}_{0.06}$ & $4.83^{0.04}_{0.12}$ & $35.61^{0.08}_{3.54}$ & $5.79^{0.24}_{0.51}$ & $1114.46^{83.78}_{288.18}$ & \dots & \dots \\
\object{HD 199260} & $3460$ &  & 2 & 2570 & 51 & $5.00^{0.08}_{0.09}$ & $8.00^{0.10}_{0.10}$ & $4.72^{0.23}_{0.21}$ & $12.67^{0.59}_{0.51}$ & $304.74^{41.28}_{28.14}$ & \dots & \dots \\
\object{HD 1581} & $3950$ &  & 3 & 4725 & 3222 & $1.32^{0.00}_{0.00}$ & $0.97^{0.00}_{0.01}$ & $5.30^{0.00}_{0.19}$ & $1.08^{0.01}_{0.01}$ & $27.26^{0.02}_{1.44}$ & $0.97^{0.06}_{0.09}$ & $806.66^{87.80}_{234.11}$ \\
\object{HD 45184} & $4420$ &  & 3 & 4751 & 312 & $0.73^{0.01}_{0.02}$ & $4.33^{0.03}_{0.04}$ & $7.33^{0.08}_{0.12}$ & $3.73^{0.07}_{0.03}$ & $27.94^{0.98}_{0.85}$ & $2.70^{0.48}_{0.17}$ & $401.98^{307.83}_{48.59}$ \\
\object{HD 154577} & $4800$ &  & 2 & 4797 & 743 & $1.01^{0.00}_{0.00}$ & $0.74^{0.01}_{0.01}$ & $7.67^{0.20}_{0.07}$ & $1.01^{0.01}_{0.01}$ & $39.11^{1.79}_{0.67}$ & \dots & \dots \\
\object{HD 43834} & $7244$ & B & 2 & 1309 & 213 & $0.72^{0.00}_{0.01}$ & $2.68^{0.01}_{0.11}$ & $34.23^{0.10}_{3.07}$ & $2.51^{0.14}_{0.18}$ & $477.80^{6.01}_{151.05}$ & \dots & \dots \\
\hline
        \hline
        \end{tabular}
        \caption{Parameters and formal errors of the 27 stars, where HD~25457 has sufficient data in HARPS and FEROS. \textit{N sines} refers to the number of periods fitted, \textit{Baseline} to the baseline the observations are spanning and \textit{\#Obs} the number of good \gls{rv} measurements. %Empty fields represent zeros ($A$) or non-fitted values ($K_i, P_i$).
        }
        \label{tab:mainresults}
        \end{sidewaystable*}
}

\begin{figure}[ht]
  \includegraphics[width=\linewidth]{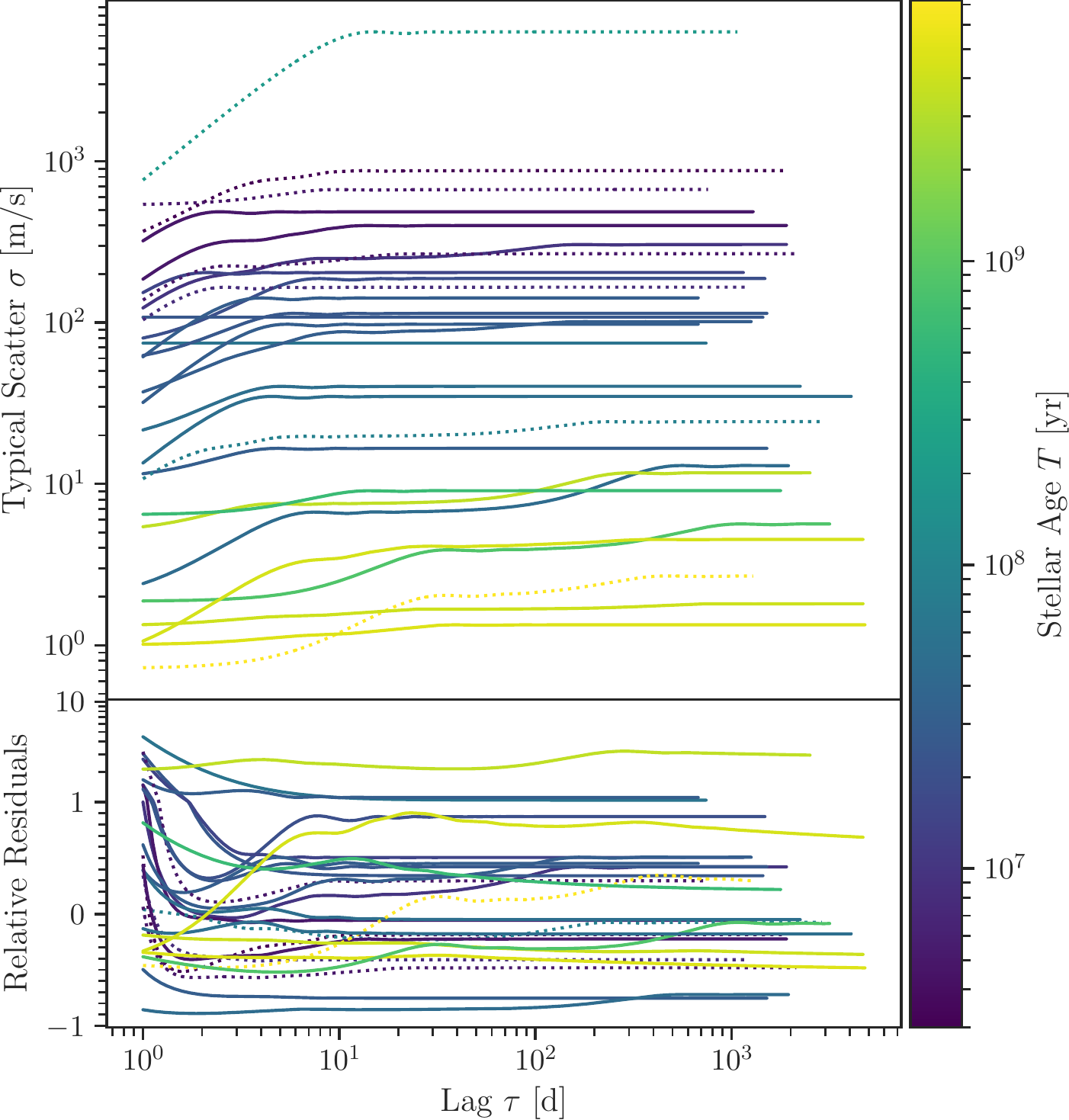}
  \caption{Top: Fitted \gls{rv} jitter of the 28 \gls{rv} datasets of the 27 stars using the procedure described in Sect. \ref{sec:modeling}. The color indicates the age of the star, showing a clear correspondence between \gls{rv} jitter and age. The \edd{dotted lines corresponds to the binaries and are not used to fit the age--lag--activity model. The outlier at the top is HD~51062, the newly identified SB. Bottom: The residuals of the above, ignoring HD~51062, from the more complex age--lag--activity model (b), divided by the model value. The model is described in Sect. \ref{sec:results}, Eq.~(\ref{eq:age_lag_actb}). We highlight the linear scale between -1 and 1 and a log-scale for values greater than 1. HD~51062 is excluded from the residuals.}
% represent the formal 1-$\sigma$ intervals, but should be considered as lower limits. You can see how the scatter is strongly age-dependant, but also it increases significantly with lag. Thus observations of stars should focus on planets with Periods $\lesssim 10 d$.
}
  \label{fig:signal_vs_timescale}
\end{figure}

We derived an analytic activity--lag--function for the 27 stars \edd{(including the 7 binaries)} that were selected by the criteria described in Sect. \ref{sec:target_sample}. The analytic function fitted to the \gls{pv} data of each star is given by Eq. (\ref{eq:multipv}). This equation describes constant noise plus up to three independent sinusoidal signals as they would show up in the \gls{pvd} in the case of infinite sampling. It has one free parameter for the constant plus two more for each signal identified. Those parameters represent the period and amplitude of the assumed underlying sinusoid. As it was integrated over the phase and we assume all phases to be covered roughly equally, no parameter for the phase or potential phase jumps is needed; see Appendix \ref{app:sinusoid}. The results of the fits are presented in Table \ref{tab:mainresults}. Figure \ref{fig:all_results_text} shows those fits graphically as well as the \gls{pv} for all stars and Fig. \ref{fig:signal_vs_timescale} then compares those to the age of the stars. With the exception of \edd{HD~51062 (highest dotted line), a clear \gls{sb} as seen in the \glspl{ccf} returned by CERES, %rotationally variable star \citep{Kiraga2012}, 
a clear correlation between age and \gls{rv} scatter $\sigma$ is found. We excluded HD~51062 as well as the other six visual binaries} from further analysis\edd{, and only use the six wide binaries as a consistency check with increased statistics below}. As shown in Table \edd{\ref{tab:mainresults}}, on average there were \edd{1.6 individual} signals with periods between \edd{2.1  and 114 days identified for those 20 stars. These signals cause an increase of the stellar noise by roughly a factor of two when comparing the variance at the smallest lags with those at the largest lags, as can be seen in the \glspl{pvd} in Fig. \ref{fig:all_results_text}.} %lead to an increase of the stellar variance by a factor of about 2 when probing longer timescales -- or lags $\tau$. 
This means that if one is looking for planetary signals, not only does the amplitude of the signal of the planet $K_\text{pl}$ decrease with $K_\text{pl} \propto P^{-1/3}$, but also the underlying noise doubles when probing months instead of days.

\begin{figure}[ht]
  \includegraphics[width=\linewidth]{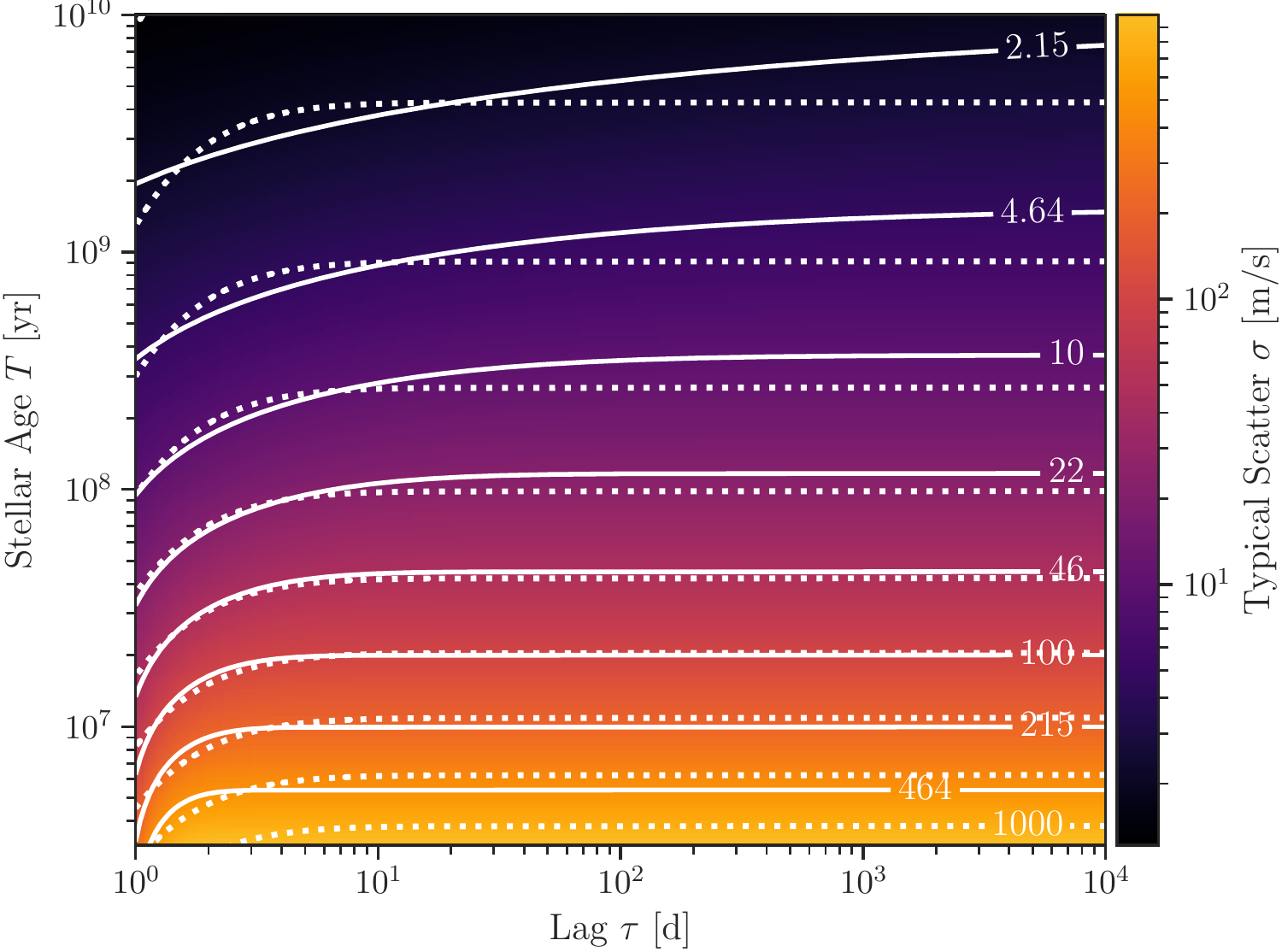}
  \caption{Fitted activity model to the \gls{pv} data excluding HD~51062. This model has been subtracted in Fig. \ref{fig:signal_vs_timescale}. The color code and white solid lines with contour levels show the more complex model (b) of Eq. (\ref{eq:age_lag_actb}).  As expected, the most important parameter is age. The increase of significant activity timescales with age can also be seen: 99\% of the final activity is reached after $\sim5$ d for a 10 Myr-old star, but only after $\sim30$ yr for a 10 Gyr-old star. The black dashed lines show the simpler model (a) of Eq. (\ref{eq:age_lag_acta}) where the timescales are forced to be the same for all ages.
  %The color code and white solid lines with contour levels show the simpler model a) of Eq. \ref{eq:age_lag_acta}, which results in a doubling of \gls{rv} jitter with $\tau$ for timescales shorter than about a month.The black dashed lines show the slightly more complex model b) of Eq. (\ref{eq:age_lag_actb}) with the same contour levels as the white ones. This model b) suggests that older stars have longer activity timescales than younger stars.
  }
  \label{fig:agelagactmodel}
\end{figure}

In order to quantify this further and to make it possible for surveys to predict the amount of \gls{rv} jitter {before} starting the observations, we fitted an empirical model, now adding a dependence on age to the model. Since the systematic errors are much larger than the formal errors on the curves, we did not account for those errors, or for the uncertainties of the ages. Instead, we reused the weights for each point derived earlier, but normalized them such that the sum of the %square root of 
weights equals one for each star. This procedure ensures that each star gets assigned the same weight, but still keeps the different weights for the individual points. \edd{The fitting is done using the python \texttt{scipy.optimize.curve\_fit} least-square fitting routine, where we fit with all three parameters (age, lag, and standard deviation) in log space, decreasing the impact of extreme values. The errors are then determined using the square root of the diagonal entries of the covariance matrix of the fit results. This can be done since the weights are scaled such that the reduced $\chi^2$ equals unity.} We used a shifted and stretched error function \edd{as our model. We chose the error function, since it asymptotically approaches different constant values in the positive and negative directions, similar to the \gls{pv} signal of a sine wave; see Eq. (\ref{eq:pvsine}). It is analytically described by
\begin{equation}\label{eq:age_lag_acta}
\text{log} \,\sigma(\hat{T}, \hat{\tau}) = \kappa_{0}\cdot \hat{T}^{\kappa_{1}} \cdot \left[ \textrm{erf}\left(\omega_{0}+\hat{\tau}\cdot \omega_1 \right)\right] -\epsilon \quad,
\end{equation}
where $\hat{T}$ and $\hat{\tau}$ are the respective decadic logarithms of the stellar age $T$ in years and lag $\tau$ in days and $\sigma$ is returned in meters per second. \edd{$\kappa_{0},\, \kappa_{1},\, \omega_{0},\, \omega_1$ and $\epsilon$} are free parameters of the model and erf is defined by $\textrm{erf}(x) := \frac{1}{\sqrt{(\pi)}} \int_{-x}^x \textrm{e}^{-t^2} \textrm{d}t$. The respective $\kappa_{i}$ and $\omega_{i}$ describe amplitude and angular frequency of the model, similar to $K$ and $2\pi / P$ in the one-dimensional case given in Eq. (\ref{eq:pvsine})}. The parameter $\omega_1$ describes how steeply the noise $\sigma$ increases with lag $\tau$, and $\epsilon$ is a simple offset since the error function goes through the origin.  We call this model (a). The fitted parameter values and errors are given in Table \ref{tab:modelparams} and the contours of the 2D function are plotted in Fig. \ref{fig:agelagactmodel} as \edd{white dotted lines}. The most striking feature is the \edd{very strong dependence} of the \gls{rv} scatter on stellar age. Although this behavior was already known qualitatively, to the authors' knowledge the dependence is quantified for the first time here. In addition, the stellar noise increases when going to longer baselines: by a factor of \edd{3.4 for stars with an age of 5 Myr years, and 1.6 for stars with an age of 5 Gyr, when comparing a baseline of 10~yr with a baseline of 1~d} . With this model, this increase happens on average such that 99\% of the maximum activity is reached \edd{after approximately 10 d}. Thus, especially for young stars, the already very high noise level is \edd{increasing from 190 m/s for a lag of 1~d to 640 m/s for a lag of 10~d or longer.}

Additionally, it would be interesting to know whether younger stars typically have longer or shorter periodic signals than older stars. To answer this, another free parameter $\delta$ was introduced, slightly changing Eq. (\ref{eq:age_lag_acta}) to
\edd{\begin{equation}\label{eq:age_lag_actb}
\text{log}\,\sigma(\hat{T}, \hat{\tau}) = \kappa_{0}\cdot \hat{T}^{\kappa_{1}} \cdot \left[ \textrm{erf}\left(\omega_{0}+\frac{\hat{\tau}\cdot \omega_{1}}{\hat{T}^{\delta}}\right)\right] -\epsilon\quad,
\end{equation}}
which we now refer to as model (b). %The same contours as for model a) are overplotted in Fig. \ref{fig:agelagactmodel} as white, solid lines.  
As shown in Table \ref{tab:modelparams}, we derive a value \edd{of $\delta=8.00\pm 0.12$. The positive value of $\delta$ means that younger stars have shorter activity timescales than older ones, as can clearly be seen in the 2D function with solid white contours shown in Fig. \ref{fig:agelagactmodel}. With this dependence, 99\% of the maximum activity will be reached after 2.5 d for stars with ages of 5 Myr and after 14 d, 800 d, and 10 kyr for stars with ages of 50 Myr, 500 Myr, and 5 Gyr, respectively. The total increase with lag is a factor of 4 for  5 Myr-old stars and a factor of 1.8 for 5 Gyr-old stars. At the same time, the absolute jitter values for a lag of 10 d decrease from  516 m/s for a 5 Myr-old star to 41 m/s, 6.7 m/s, and 1.9 m/s for stars with ages of  50 Myr, 500 Myr, and 5 Gyr, respectively.}
%However, since the error is larger than the value, we cannot confirm this trend as significant.
%How fast this increase occurs, depends on the star's age. For young stars ($\lesssim$ 20 Myr), the noise plateaus already after about a week. For old stars ($\gtrsim$ 500 Myr), the increase only settles after years or even decades. This hints that long term \gls{rv} noise inducing signals are relatively stronger in old, inactive stars, when compared to their shorter term variations. Nonetheless, only for stars of very similar age, the younger star can be less noisy when probed on much shorter timescales.
\begin{table}
        \begin{tabular}{lcccccc}
        \hline \hline
        Mod.&$\kappa_{0}$ &$\kappa_{1}$ &$\omega_{0}$ &$\omega_{1}$ & $\delta$&$\epsilon$\\
%       &\edd{[m/s]}&&&&&\edd{[m/s]}\\
        \hline
%       b)&$-11.48$ & $-0.86$ & $1.15$ & $58982$ & $5.35$ & $2.29$\\
%       &$\pm0.24$&$\pm0.01$&$\pm0.01 $&$\pm26193 $&$ \pm0.22$&$\pm0.05 $\\
    a)&$454$ & $-2.463$ & $1.077$ & $1.164$ &\dots & $1.382$\\
        &$\pm31$&$\pm0.040$&$\pm0.007 $&$\pm0.038 $&\dots &$\pm0.041 $\\
        b)&$312$& $-2.294$& $1.015$&12.8M&$8.00$& $1.261$\\
         &$\pm 20$&$\pm 0.040$&$\pm 0.008$&$\pm3.7$M&$\pm 0.12$&$\pm 0.041$\\
\hline
        \end{tabular}
        \caption{Least-square fit results of the simple model (a) from Eq. (\ref{eq:age_lag_acta}) and the slightly more complex model (b) described in Eq. (\ref{eq:age_lag_actb}) to the data of \edd{the 20 putatively single stars.} The first row gives the parameter values, the second the corresponding formal errors.}
        \label{tab:modelparams}
\end{table}

 
\begin{figure}[ht]
  \includegraphics[width=\linewidth]{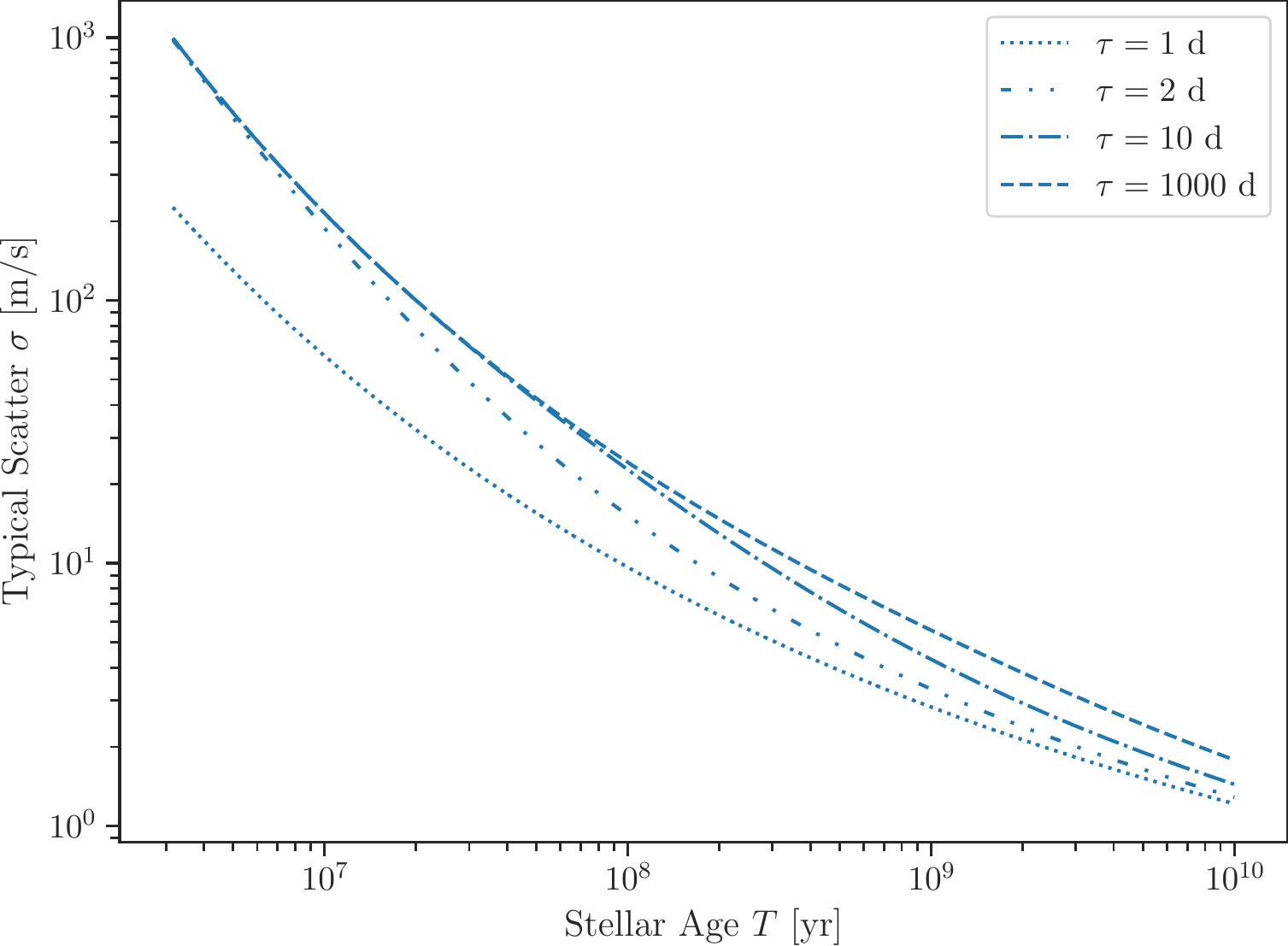}
  \caption{\edd{Four slices for different lags through the activity model (b) presented in Fig. \ref{fig:agelagactmodel} with solid lines. The slices show the strong dependance on age, but also on lag, in particular for the older stars.}}
  \label{fig:agelagactmodel1d}
\end{figure}
%--------------------------------------
\section{Discussion}\label{sec:discussion}
Without making use of any stellar models, we were able to determine the \gls{rv} jitter as function of lag for 27 stars and to describe it with an analytic function for all of those using \gls{pv}. However, since the assumption of statistical independence is violated in the pooled points, the F-test used to determine the number of sinusoidal signals identified is strictly speaking not applicable. Thus, even though the results appear convincing, one cannot put numbers on the significance of an identified signal, one of the original plans to characterize stars even further. Consequently, the errors determined using bootstrapping need to be considered as lower limits because of this lack of statistical independence. For the future, a \gls{mc} simulation on the original data with scaled error bars would perhaps return more realistic errors. But since we neither make use of the number of signals identified nor of the errors, we did not pursue this further. 

\edd{This lack of statistical independence has the largest influence when little and clustered data is present. This is often the case in the HARPS data taken after the intervention, resulting in a horizontal line of pooled data, as seen for example in HD~30495 in Fig. \ref{fig:all_results_appendix}. This is because often only a few or even one `box' has at least two data points, allowing calculation of the variance. In the case of HD~30495, multiple data points were taken in one night, with the shortest time period between two nights of data capture being 44~d. Thus, the \gls{pv} only changes for a lag of 44~d or larger, sometimes resulting in a jump in the \gls{pvd}. Fortunately, fewer data points are used in these cases, down-weighting the influence of these cases. Also, the selection criteria presented in Sect. \ref{sec:target_sample} try to limit the occurrence of these phenomena.}

\edd{For stellar activity there are three main characteristic timescales involved \citep[e.g.,][]{Borgniet2015}\edl{: First, the stellar rotation causing rotational modulation of active regions and their effects with typical timescales of days to a few weeks. Second, the reconfiguration of active regions with typical timescales of several weeks. And third, the stellar activity cycle with typical timescales of several years or decades.} Of those three,  \edl{the second} is the least periodic \edd{and has a rather unsharp timescale}; furthermore, it overlaps \edl{with the stellar rotation timescale}, meaning that the two are difficult to disentangle; see for example \citet{Giles2017}. Many stars show an increase of the \gls{rv} jitter up to a few days or weeks followed by a plateau. This is probably due to the stellar rotation and reconfiguration of the active regions. In addition, especially the older, more quiet stars show another strong increase with a lag of a few years. This is probably related to the stellar activity cycle. However, since this increase is often close to the probed baseline, we might often only probe a part of this additional noise term.
}

\edd{For very active stars, the white noise term $A$ of Eq.~\ref{eq:multipv}  is often \edd{consistent with zero, taking the error bars into account}. This probably means that the periodic signal dominates the fit and therefore $A$ cannot be determined very well numerically instead of truly being zero.}

Since the ages of all stars are known, we could determine an empirical but still analytic model of the age--lag--activity relation \edd{based on the 20 presumably single stars}. This model shows that the typical \gls{rv} jitter of a star depends \edd{strongly }%more strongly than exponentially 
on the age of the star, \edd{ranging from 516 m/s for a 5 Myr-old star down to 2.5 m/s for a 5 Gyr-old star -- when probing lags of 10 yr.} %This makes the age a crucial parameter for the presence of \gls{rv} jitter. The strength of the increase of the \gls{rv} jitter with lag depends on the age: The maximum stellar RV jitter of 3 million year old stars is 820 m/s, and reduces to 199 m/s,  23 m/s and 4.8 m/s and 1.5 m/s  for stars with ages of 10 Myr, 100 Myr, 1 Gyr and 10 Gyr,  respectively. 
When considering a timescale of 1 d only, the \gls{rv} jitter is smaller by roughly a factor of \edd{4 for stars with ages of 5 Myr and a factor of 1.7 for stars with an age of 5 Gyr. The rate by which this jitter then increases with lag strongly depends on the age of the stars, where according to the model younger stars reach their maximum in a few days and older stars take up to thousands of years.} 
%99\% of the maximum \gls{rv} jitter will be reached after \edd{3 d, 5 d, 19 d, 225 d and 30 yr for stars with an age of 3 Myr, 10 Myr, 100 Myr, 1 Gyr and 10 Gyr}, respectively. 
We note that the maximum baseline of our measurements is about 13 yr, meaning that we cannot say anything about the contribution of activity cycles that operate on even longer timescales. \edd{The ages of the single stars probed reach from 5 Myr to 5 Gyr and cover spectral types from  F6 to M0; see Table~\ref{tab:stellarproperties} and Fig.~\ref{fig:agedist_all}. By construction, each star is assigned the same weight, and since 14 of the 20 stars are younger than 100 Myr, the result is skewed towards relatively young stars and thus is also most robust for those young stars. We also note that the individual scatter of $\sigma_\text{p}$ of the stars is about a factor of two, as can be seen in the residuals of Fig.~\ref{fig:agelagactmodel}. This might hint towards another dependence of the activity on spectral type or metallicity. However, more data are needed to analyze this dependence.}

\edd{To test our model with increased statistics, we derived the age--lag relationship of model (b) with the six additional wide ($\ge 27$ AU projected separations) binary systems, that is, all stars except the \gls{sb} HD~51062. Since the activity of these stars is on average slightly lower than that of the similar-aged single stars (see Fig. \ref{fig:signal_vs_timescale}) the \gls{rv} data for these stars are likely not dominated by the binary companions, justifying this approach. The derived values are $\kappa_0 = 98.4 \pm 4.1,\, \kappa_1 = -1.544\pm 0.029,\,\omega_0 = 1.1612\pm 0.0071,\, \omega_1 = (3.17\pm 1.56)\cdot 10^6,\, \delta = 7.36 \pm 0.25 \text{ and } \epsilon = 2.471 \pm 0.066\, \text{m/s}$. Despite individual values changing more than the formal errors (e.g., $\kappa_0$ from 312 to 98), the typical scatter $\sigma$  does not change by more than 20\% in the probed parameter space (ages from 5 Myr $-$ 10 Gyr, lags from 1 d $-\, 10^4$ d). Also the trend of longer activity timescales for older stars, quantified by the $\delta$ parameters, remains of the same order and clearly positive. The major change is a decrease of the \gls{rv} scatter by slightly less than 20\% for 5 Myr-old stars and by 10\% for 5 Gyr-old stars. It remains the same within a few percent for stars of intermediate ages ($\sim$50 Myr -- 1 Gyr). Therefore, this test strengthens our confidence in the robustness of our model.}

%\edd{Thus having analyzed 26 individual stars, we believe that our results from Table \ref{tab:mainresults} can be used to predict the stellar activity induced \gls{rv} variations for stars with a given age and a given timescales with a precsion of about a factor of two, compare the relative residuals in Fig. \ref{fig:signal_vs_timescale}.}

%It roughly triples for stars with the age of $10^{6.5}$ years from 331 m/s to 1040 m/s, and increases by a factor of 1.8 from 2.2 m/s to 4.1 m/s for 1 Gyr year old stars when probing months instead of a few days. For stars even older, the limit is usually determined by instrumental precision; the model suggests 0.7 m/s for a lag of 1 d and 1.2 m/s for a lag of 3 yr, which is of the order of the HARPS precision of 1 m/s.\xxx

%Keeping the timescale of this increase to be the same for all stars (model a)), we find that 99\% of the maximum \gls{rv} jittter will be reached with a lag of 20 days, whereas the relative increase and the maximum values of the \gls{rv} jitter do not change significantly. %Allowing this timescale to vary with age hints towards significant noise contribution even on the timescale of years for stars older than about 1 Gyr and a timescale of just a few days for very young stars. However, the corresponding fit parameter's formal error is only 2 times smaller than the value itself, rendering the result in question.

This means that for \gls{rv} exoplanet-hunting surveys using state of the art instruments for example, one should be aware that for stars younger than a few hundred million years the limit is set by the stellar jitter and not by instrumental precision and the age is a crucial parameter. For example, the RV\,SPY survey excludes stars younger than about 5 Myr although its goal is  to find planets around young stars. Additionally, RV\,SPY focuses on searching for hot Jupiters, where the stellar jitter is slightly smaller and the signal of the planet is larger than for longer periodic planets (Zakhozhay et al., in prep.).

\section{Conclusions}\label{sec:conclusions}
Here, we show that the \gls{pv} method can be used to model and determine stellar activity timescales and amplitudes. We applied PV to 28 data sets of 27 different stars. We find an empirical relation between \gls{rv} scatter, stellar age, and lag $\tau$. %\edd{using 21 of those data sets and strengthening this result using another 6 of wide binary systems}. 
We found a \edd{very strong} dependence of \gls{rv} jitter on age. Further, the \gls{rv} scatter roughly doubles with the lag $\tau$ when probing over timescales of months instead of a few days.

This relation is not only important for stellar modeling, but also for developing an observing strategy for \gls{rv} exoplanet surveys, especially if young stars are involved\edd{, such as predicting the \gls{rv} jitter for young K/G dwarfs common in the TESS survey. Also,} in searches for hot Jupiters, dense sampling is approximately twice as sensitive as long-term random sampling. One survey making use of the findings in this paper is the RV\,SPY survey with FEROS, looking for hot Jupiters around young stars (Zakhozhay et al., in prep.).

\begin{acknowledgements}
This research has made use of the NASA Exoplanet Archive, which is operated by the California Institute of Technology, under contract with the National Aeronautics and Space Administration under the Exoplanet Exploration Program.\\
This research made use of Astropy (\url{http://www.astropy.org}) a community-developed core Python package for Astronomy \citep{astropy:2013, astropy:2018}.\\
This research has made use of the SIMBAD database,
operated at CDS, Strasbourg, France \\
Thanks also to Diana Kossakowski for fruitful discussions about Gaussian processes and its limitations on this application and to Peter Markowski for his help on statistical issues.\\
\edd{We would also like to thank the anonymous referee for his or her helpful comments and ideas, which helped to improve the paper.}
\end{acknowledgements}

% WARNING
%-------------------------------------------------------------------
% Please note that we have included the references to the file aa.dem in
% order to compile it, but we ask you to:
%
% - use BibTeX with the regular commands:
%   \bibliographystyle{aa} % style aa.bst
%   \bibliography{Yourfile} % your references Yourfile.bib
%
% - join the .bib files when you upload your source files
%-------------------------------------------------------------------
\bibliographystyle{aa}
\bibliography{bibliography}

\appendix
\section{Analytic description of the PV}
The variance of a data set $y_i = 1,\dots,n$ is given by
\begin{equation}
  \sigma^2(n)=\frac{1}{n-1}\sum_{i=1}^n(y_i-\overline{y})^2
,\end{equation}
where $\overline{y}=\frac{1}{n}\sum_{i=1}^n y_i$. For sufficiently large $n$, we have
\begin{equation}
  \label{eq:sinusapprox}
  \sigma^2\approx \frac{1}{n}\sum_{i=1}^n(y_i-\overline{y})^2=\overline{y^2}-\overline{y}^2
,\end{equation}
where $\overline{y^2}=\frac1n\sum_{i=1}^ny_i^2$.

In the limit of equidistant and infinitely dense sampling of the data, we can replace the sums by integrals;  Eq. \ref{eq:sinusapprox} then becomes
\begin{equation}\label{eq:pvinf}
  \sigma^2(\tau) = \frac1\tau \int_0^\tau y^2(t) \,\textrm{d}t-\left( \frac1\tau \int_0^\tau y(t) \,\textrm{d}t \right)^2
,\end{equation}
where $\tau$ is the timescale over which the variance is to be evaluated.

As we see below, for periodic functions the shape of $\sigma^2(\tau)$ depends strongly on the phase of the periodic function. In practice however, the phase of a signal is often sampled repeatedly in a random fashion, \edd{thus also averaging over potential variations in the phase.} Therefore, we consider the phase-averaged variance, with $\delta \in [0, 2\pi)$ the phase of the signal $y(\delta)$
\begin{equation}\label{eq:phaseav}
  \overline{\sigma^2}(\tau) = \frac{1}{2\pi}\int_0^{2\pi}\sigma^2(\tau, \delta) \,\textrm{d}\delta
.\end{equation}

We note that we take averages over the signal $\overline{y}$ with respect to the selected time scale $\tau$, whereas we take the average over the variance $\overline{\sigma^2}$ with respect to the phase of the signal $\delta$.

In the analytic case (infinitely dense sampling) the \gls{pv} of a data set is given by inserting Eq. \ref{eq:pvinf} into Eq. \ref{eq:phaseav}.  In reality one might see effects of transient oscillations if the signal is not sampled densely at all phases. To minimize the influence of those effects, the selection criteria for the sampling described in Sect. \ref{sec:target_sample} were applied.

\section{Analytic PV of sinusoids}\label{app:sinusoid}
As an example we consider the \gls{pv} of a sine wave:
\begin{equation}\label{eq:sine}
  y(t) = K \sin \left( \frac{2\pi}{P}t-\delta \right)
,\end{equation}
where $K$ is the (semi-) amplitude, $P$ is the period, and $\delta\in [0, 2\pi)$ the phase.

Evaluating Eq. \ref{eq:pvinf} for the sine function of Eq. \ref{eq:sine} yields the scaled \edd{variance
\begin{align}\label{eq:b2}  \begin{split}
    \Omega^2(\theta) = \frac12 &- \frac{\sin(2\theta-2\delta)+\sin (2\delta)}{4\theta}\\
    &- \left[ \frac{\cos \delta - \cos(\theta -\delta)}{\theta} \right]^2,
  \end{split} \end{align}
where $\Omega^2 \coloneqq \sigma^2/K^2$ and $\theta = 2\pi\tau/P$.
}
Inserting Eq. \ref{eq:b2} into Eq. \ref{eq:phaseav} yields the normalized, analytic \gls{pv} of a \edd{sine wave
\begin{equation}
  \overline{\Omega^2}(\theta) = \frac12+\frac{\cos\theta -1}{\theta^2}
,\end{equation}}
or
\begin{equation}\label{eq:pvsine}
  \overline{\sigma^2}(\tau) = K^2\left[ \frac12 + \frac{\cos (2\pi\tau/P)-1}{(2\pi\tau/P)^2} \right]
.\end{equation}
We have $\lim\limits_{\tau \rightarrow 0}{\overline{\sigma^2}(\tau)} = 0$ and $\overline{\sigma^2}(\tau)=K^2/2$ for $\tau = P$ and also for $\tau \gg P$.

\section{Radial-velocity data}
See next page.
\newpage
\begin{figure*}[htb]
  \includegraphics[width=\textwidth]{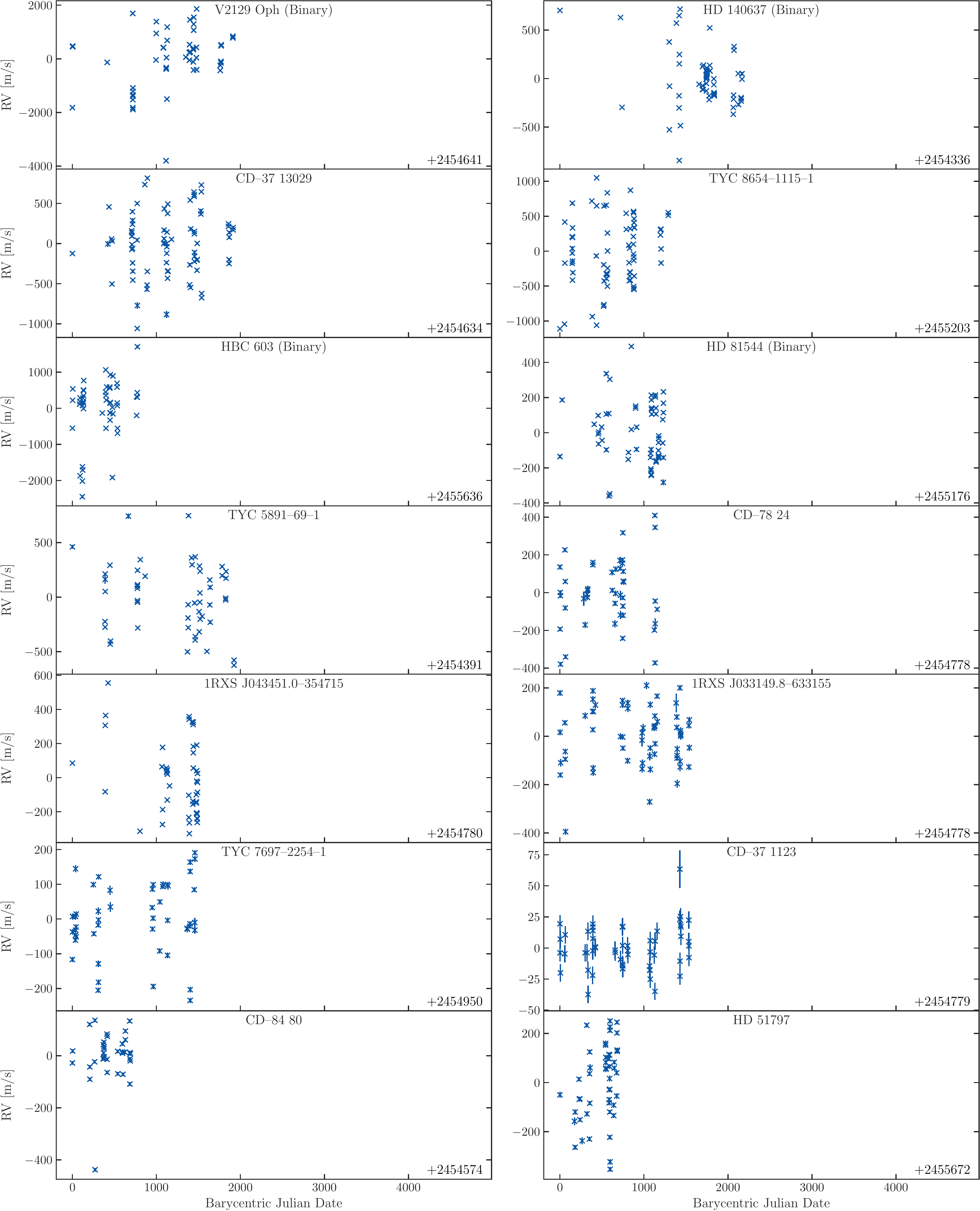}
  \caption{Radial velocity data for all 27 stars after removing the bad data as described in Sect. \ref{sec:target_sample} and subtracting the mean of each data set. Green symbols denote HARPS and blue symbols FEROS data.  The offset of the Julian Date on the x-axis is given in
the bottom right of each plot. The green vertical lines mark the fiber change of HARPS where the data sets were split. One can see how some data show clear jumps there while others do not. Jumps in the \gls{pvd} in Fig. \ref{fig:all_results_text} as in HD~51062 or HD~25457 can be explained by clustered (e.g., HD~25457) or relatively sparse (e.g., HD~51062) data.}
  \label{fig:all_rvdata}
\end{figure*}
\begin{figure*}[htb]\ContinuedFloat
  \includegraphics[width=\textwidth]{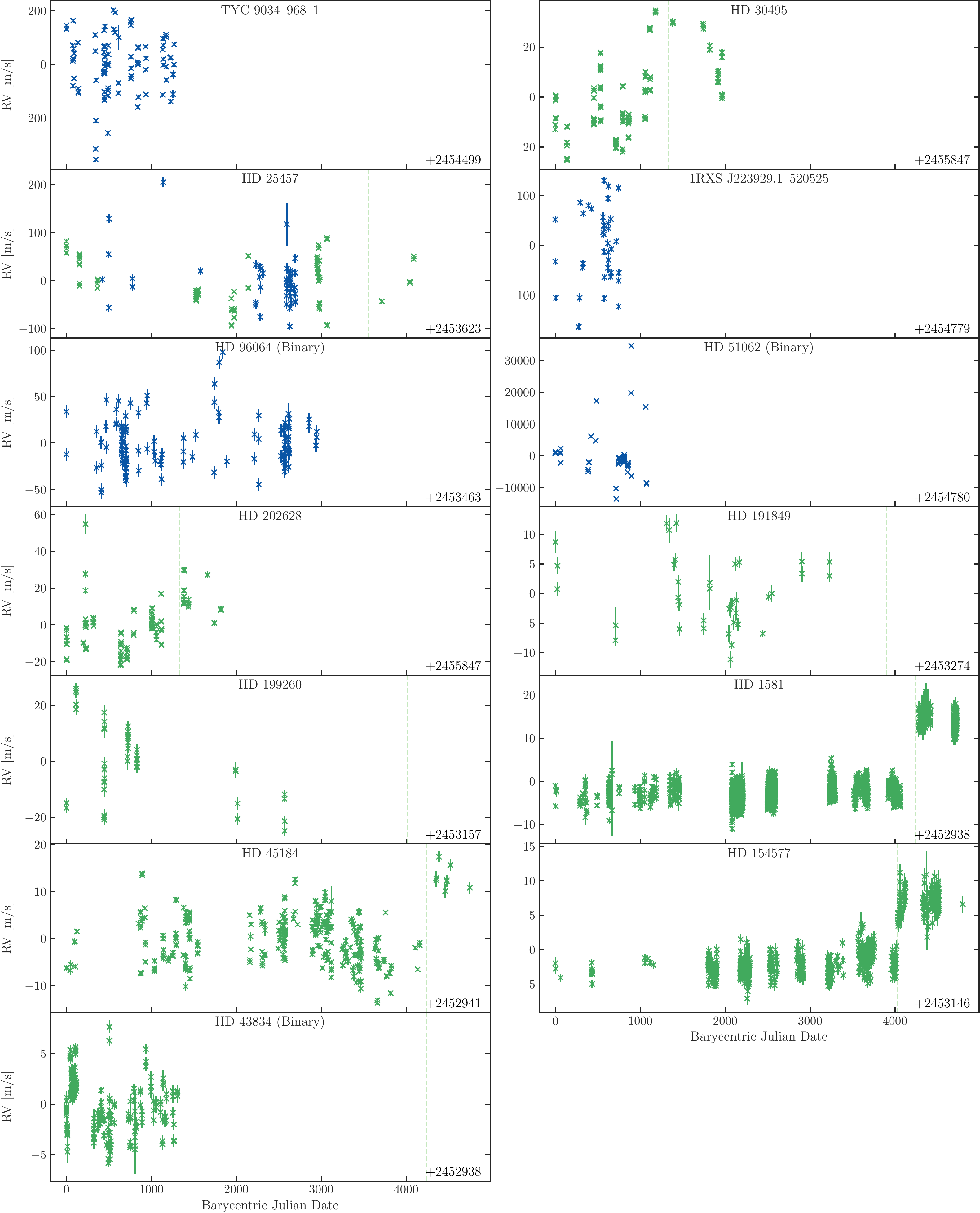}
  \caption{Continued.}
  \label{fig:all_rvdata2}
\end{figure*}

\FloatBarrier
\section{Pooled variance diagrams}
\bigskip
\begin{minipage}[c]{\textwidth}
  \centering
  \includegraphics[width=\textwidth]{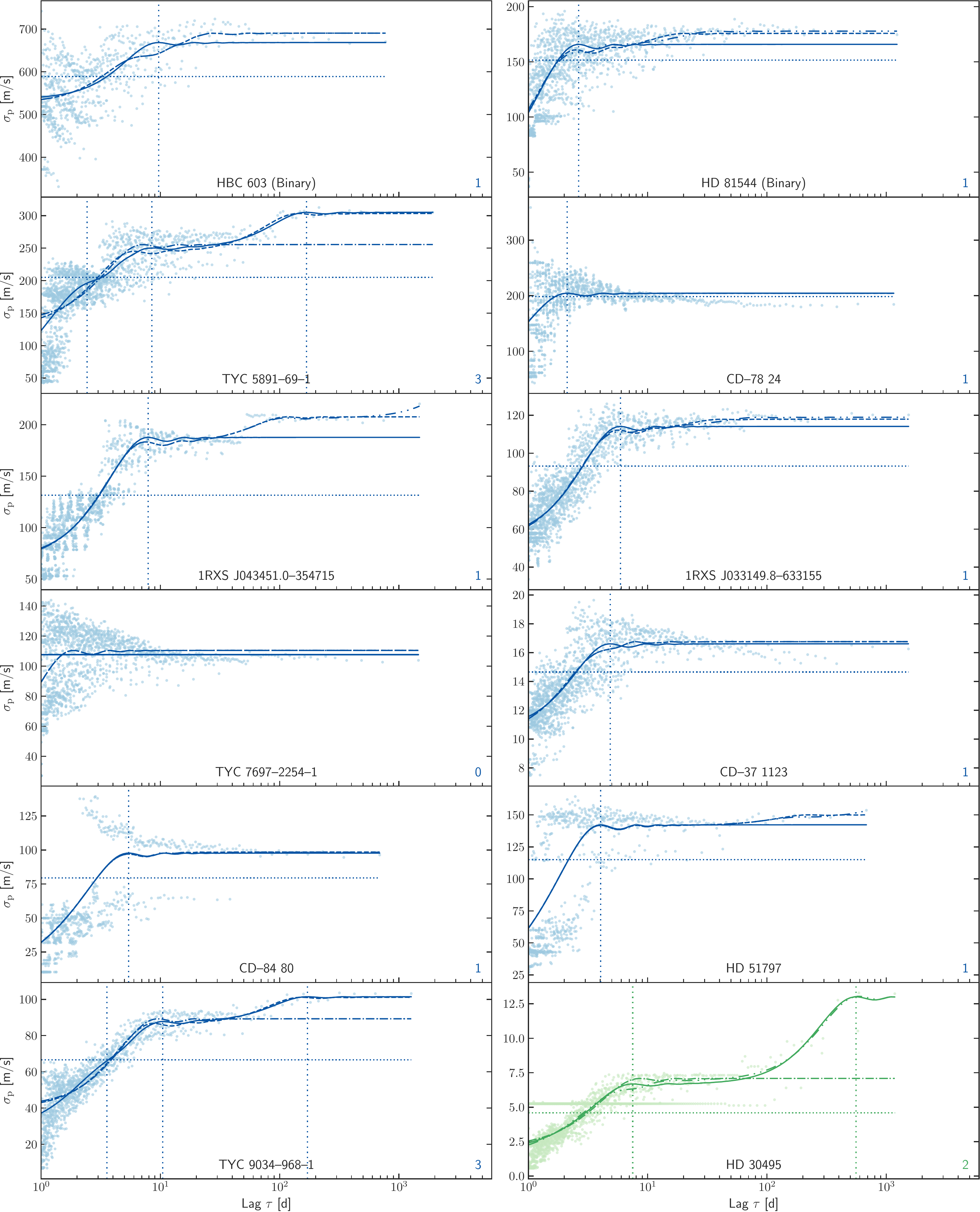}
  \captionof{figure}{Continued from Fig. \ref{fig:all_results_text}.}
  \label{fig:all_results_appendix}
\end{minipage}
\clearpage
\noindent
\addtocounter{figure}{-1}
\begin{minipage}[c]{\textwidth}
  \centering
  \includegraphics[width=\textwidth]{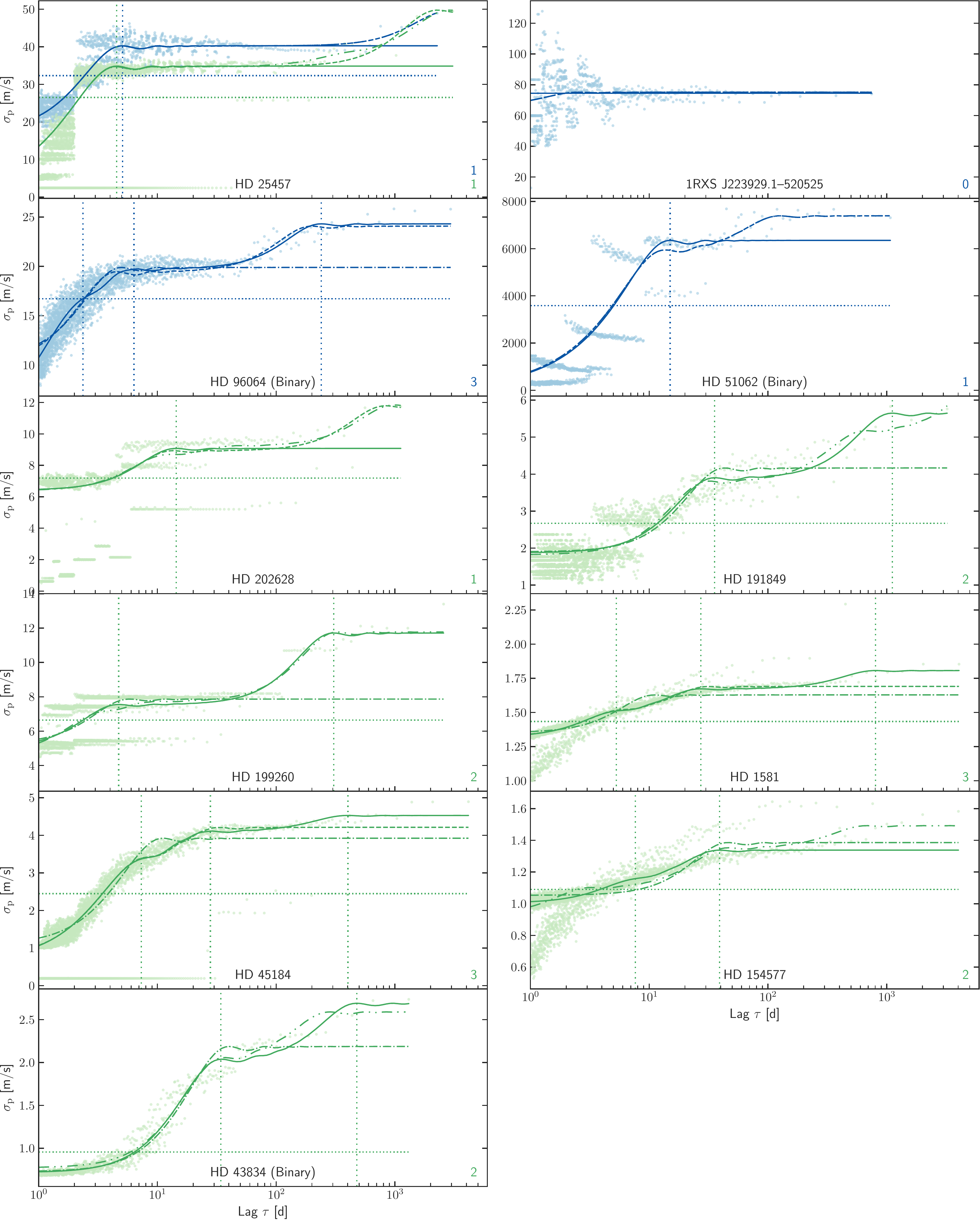}
  \captionof{figure}{Continued.}
  \label{fig:all_results_appendix2}
\end{minipage}
\clearpage
\section{Pooled variance error estimate}
\bigskip
\begin{minipage}[c]{\textwidth}
\centering
  \begin{overpic}[width=\textwidth]{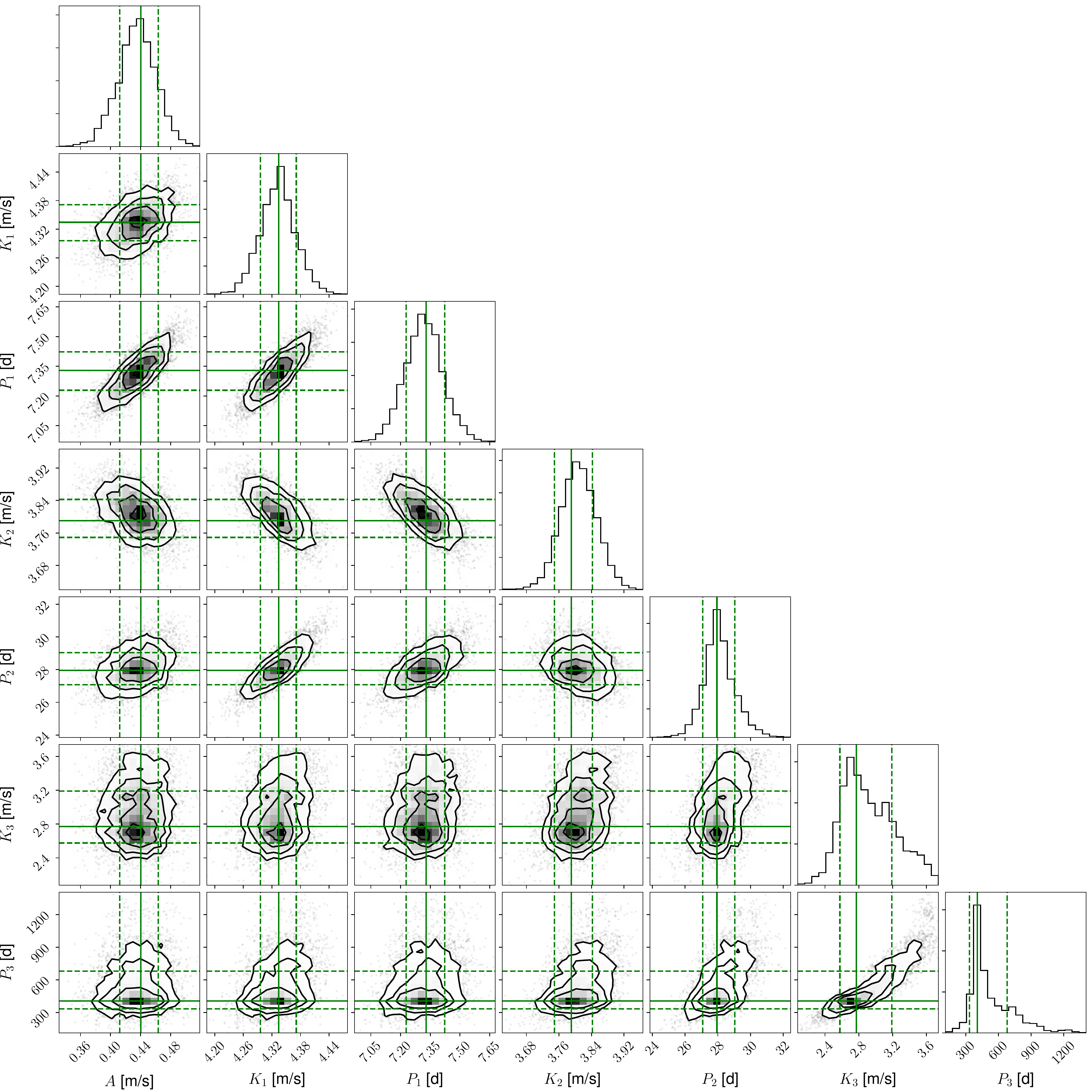}
  		\put(425, 632){\includegraphics[width=0.575\textwidth]{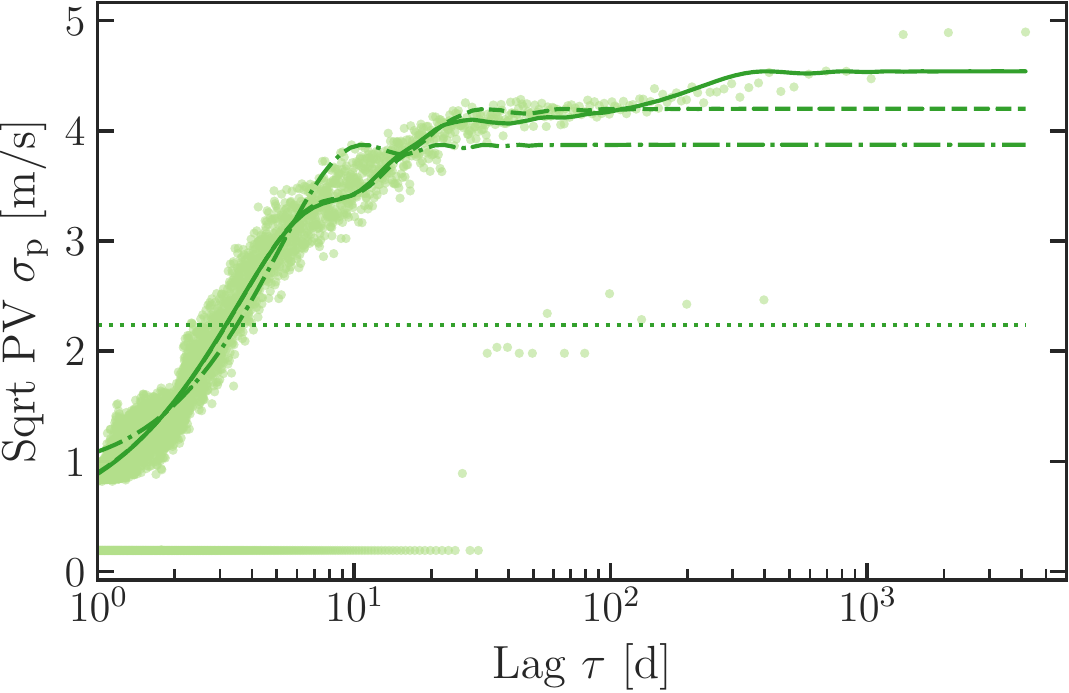}}
  		\put(260, 980){\large{HD 45184}}
    \end{overpic}
  \captionof{figure}{Bottom left: Results of the bootstrapping procedure shown as an example for HD~45184. Shown is a corner plot using the \citet{Foreman2016} python package. $A$ denotes the offset wheres $K_i$ and $P_i$ show the amplitudes and periods of the $i$th signal, as defined in equation \ref{eq:multipv}. The solid green bars denote the fit to the original data, whereas the dashed lines denote the $1\,\sigma$-confidence levels. In the case of HD~45184, three signals were identified as significant by the F-test, which are shown here. Top right: PV plot: The points mark the results of the \gls{pv} and the curves the fits with different numbers of sinusoidal signals modeled: Dotted: zero (constant), dash-dotted: one sinusoid, dashed: two sinusoids, solid: three sinusoids (best fit). The flat line of points at the bottom are due to the sparse sampling after the HARPS intervention, where only two times two observations are taken within less than 20 days, and those were respectively taken on the same nights; see Fig. \ref{fig:all_rvdata}.% But since they are down-weighted anyway and to be consistent, we keep them.
  %The same plot for all stars is seen in Fig. \ref{fig:all_results}.
  }
  \label{fig:bootstrcornerplot}
\end{minipage}
\end{document}